\documentclass[12pt,a4paper]{article}

\usepackage{url}
\usepackage{amsfonts}
\usepackage{amsmath,amssymb,color}
\usepackage{graphicx}
\usepackage{enumitem}
\usepackage{multirow}
\usepackage{bm}
\usepackage{natbib}

\usepackage{enumitem}
\newlist{constraints}{enumerate}{1}
\setlist[constraints]{
  label={[C\arabic*]},    
  ref={[C\arabic*]},      
  leftmargin=*,
  align=left,
  itemsep=4pt,            
  parsep=0pt
}

\usepackage{verbatim}
\usepackage{float}
\usepackage[toc,page]{appendix}
\usepackage{setspace,fullpage}
\usepackage{setspace}
\usepackage{adjustbox}
\usepackage{rotating}
\usepackage{subcaption}
\usepackage{amsthm}
\usepackage{booktabs,dcolumn,caption}
\usepackage{accents}
\usepackage{algpseudocode}
\usepackage{algorithm}
\usepackage{xr}

\usepackage{tikz}
\usetikzlibrary{shapes.geometric, arrows}

\tikzstyle{startstop} = [rectangle, rounded corners, 
minimum width=1.5cm, 
minimum height=1cm,
text centered, 
draw=black]

\tikzstyle{io} = [trapezium, 
trapezium stretches=true, 
trapezium left angle=70, 
trapezium right angle=110, 
minimum width=3cm, 
minimum height=1cm, text centered, 
draw=black, fill=blue!30]

\tikzstyle{process} = [rectangle, 
minimum width=3cm, 
minimum height=1cm, 
text centered, 
text width=3cm, 
draw=black, 
fill=orange!30]

\tikzstyle{decision} = [diamond, 
minimum width=2cm, 
minimum height=1cm, 
text centered, 
draw=black]
\tikzstyle{arrow} = [thick,->,>=stealth]

\usepackage{array}
\newcolumntype{L}[1]{>{\raggedright\let\newline\\\arraybackslash\hspace{0pt}}p{#1}}
\newcolumntype{C}[1]{>{\centering\let\newline\\\arraybackslash\hspace{0pt}}p{#1}}
\newcolumntype{R}[1]{>{\raggedleft\let\newline\\\arraybackslash\hspace{0pt}}p{#1}}

\ifdefined\TimesFont 
\usepackage{times} 
\fi

\ifdefined\ParSkip 
\usepackage{parskip} 
\fi

\newtheorem{theorem}{Theorem}

\newtheorem{assumption}{Assumption}

\newtheorem{corollary}{Corollary}

\newtheorem{example}{Example}

\newtheorem{lemma}{Lemma}

\newtheorem{proposition}{Proposition}
\newtheorem{remark}{Remark}

\makeatletter
\newcommand*\rel@kern[1]{\kern#1\dimexpr\macc@kerna}
\newcommand*\widebar[1]{%
  \begingroup
  \def\mathaccent##1##2{%
    \rel@kern{0.8}%
    \overline{\rel@kern{-0.8}\macc@nucleus\rel@kern{0.2}}%
    \rel@kern{-0.2}%
  }%
  \macc@depth\@ne
  \let\math@bgroup\@empty \let\math@egroup\macc@set@skewchar
  \mathsurround\z@ \frozen@everymath{\mathgroup\macc@group\relax}%
  \macc@set@skewchar\relax
  \let\mathaccentV\macc@nested@a
  \macc@nested@a\relax111{#1}%
  \endgroup
}
\makeatother

\DeclareMathOperator*{\argmax}{argmax}

\DeclareMathOperator{\tr}{tr}

\DeclareMathOperator{\ve}{vec}

\newcommand{\aaa}{{\boldsymbol \alpha}}
\newcommand{\bbb}{{\boldsymbol \beta}}

\newcommand{\LLLL}{\boldsymbol \Lambda}
\newcommand{\ttt}{\boldsymbol \theta}
\newcommand{\TTT}{\boldsymbol \Theta} 
\newcommand{\SG}{\boldsymbol \Sigma}

\newcommand{\ph}{\boldsymbol \phi}
\newcommand{\PH}{\boldsymbol \Phi}

\newcommand{\A}{\mathbf A}
\newcommand{\AAA}{\mathcal A}

\newcommand{\BBB}{\mathcal B}

\newcommand{\CC}{\mathbf C}

\newcommand{\DD}{\mathbf D}

\newcommand{\ee}{\mbox{$\mathbf e$}}
\newcommand{\EEE}{\mathbb E}
\newcommand{\EEEE}{\mathcal E}

\newcommand{\FF}{\mbox{$\mathbf F$}}

\newcommand{\JJ}{\mathbf J}

\newcommand{\MM}{\mathbf M}
\newcommand{\MMM}{\mathcal M}
\newcommand{\XX}{\mathbf X}

\newcommand{\YYY}{\mathcal Y}

\newcommand{\HH}{\mathbf H}

\newcommand{\hh}{\mathbf h}
\newcommand{\II}{\mathbf I}

\newcommand{\PP}{\mathbf P}

\newcommand{\R}{\mathbb R}
\newcommand{\RR}{\mathbf R}

\newcommand{\SSSS}{\mathcal S}

\newcommand{\uu}{\mathbf u}
\newcommand{\UU}{\mathbf U}
\newcommand{\VV}{\mathbf V}
\newcommand{\vv}{\mathbf v}

\newcommand{\ww}{\mathbf w}

\newcommand{\XXX}{\mathcal X}

\newcommand{\Z}{\mathbf Z}

\newcommand{\QQ}{\mathbf Q}

\newcommand{\1}{\uppercase\expandafter{\romannumeral1}}
\newcommand{\2}{\uppercase\expandafter{\romannumeral2}}

\usepackage{hyperref} 
\usepackage{tcolorbox}
\hypersetup{
    colorlinks=true,
    linkcolor=[HTML]{297DCF},
    citecolor=[HTML]{297DCF},
    filecolor=[HTML]{297DCF},
    urlcolor=[HTML]{297DCF},
    bookmarks=true,            
    bookmarksopen=true,        
}

\usepackage{stmaryrd}
\usepackage{booktabs}
\usepackage{caption}
\usepackage{subcaption}
\usepackage{multirow}
\usepackage{pdflscape}
\usepackage{float}  

\def\ParSkip{} 

\algrenewcommand\algorithmicensure{\textbf{Output:}}

\title{Unfolding and Fusion: Debiased Inference for Generalized Multilayer Latent Space Models}
\author{Zhaozhe Liu$^{\dag}$, Gongjun Xu$^\ddag$ and Haoran Zhang$^{\dag}$\\ [10pt]
	$^\dag$Department of Statistics and Data Science \\
	Southern University of Science and Technology 
	\and
	$^\ddag$ Department of Statistics \\
	University of Michigan 
}
\date{}

\begin{document}
\maketitle


\doublespacing


\begin{abstract}
Multilayer networks have become increasingly ubiquitous across diverse scientific fields, yet their inferential theory remains underdeveloped. We propose a flexible latent space model for directed multilayer networks with various edge types, where each node has sender and receiver latent positions capturing its sending and receiving behaviors, together with layer-varying out- and in-degree parameters and layer-specific connection matrices. Through nonlinear link functions, the model induces a Tucker low-rank tensor structure in which the structural latent spaces are intertwined with the degree-effect subspaces, creating challenges for estimation and inference. To address these challenges, we develop a novel unfolding and fusion method, coupled with two-sided centering to partial out degree effects. We establish consistency and asymptotic normality for the latent-position estimators. For the connection matrices, we uncover a non-negligible second-order bias induced by fusing two separately estimated unfolding components, derive its explicit representation, and construct a feasible plug-in correction. The resulting debiased estimator is asymptotically normal. Together, these results enable confidence regions for latent positions and tests of whether two network layers share the same structure. Extensive simulation studies validate the proposed method, and a real-world data analysis demonstrates its practical utility.
\end{abstract}



\section{Introduction}

Multilayer networks serve as a powerful representation for relational data, in which nodes correspond to entities and edges across different layers capture multiple types of relationships among them. Such networks have become increasingly prevalent in a variety of real-world applications, including biological networks \citep{liuRobustnessLethalityMultilayer2020, nunez-carpinteroRareDiseaseResearch2024}, international trade networks \citep{a.alvesUnfoldingComplexityGlobal2018, renBridgingNestednessEconomic2020, jing2021community}, and social networks \citep{baggioMultiplexSocialEcological2016, dickison2016multilayer}.

Over the last decade, many models and methods have been developed to facilitate the analysis of multilayer networks. For example, community detection for multilayer networks has been extensively studied, based on the multilayer stochastic block model and its variants \citep{paul2016consistent, barbillon2017stochastic, wilson2017community, yuan2021community, jing2021community}. Various methods have been developed for community detection, including the spectral methods \citep{bhattacharyya2017spectral, xie2024bias}, maximum likelihood estimates \citep{han2015consistent, paul2016consistent, yuan2021community}, least square estimates \citep{lei2020consistent}, and tensor decomposition \citep{jing2021community}. Besides the multilayer stochastic block model, multilayer latent space models \citep{gollini2016joint, salter2017latent, d2019latent} are more flexible models for multilayer networks, which originate from the latent space model for a single network \citep{hoff2002latent}. The common idea behind different multilayer latent space modeling is that, besides node-wise latent positions, there are additional layer-wise structure parameters, which correspond either to degree heterogeneity across layers \citep{gollini2016joint, he2025semiparametric} or to layer structures, such as layer-wise scalars, latent positions, or connection matrices \citep{d2019latent, zhang2020flexible, macdonald2022latent, zhang2025efficient}. Average consistency results of the maximum likelihood estimators for latent positions have been established \citep{zhang2020flexible, macdonald2022latent, zhang2025efficient}. Recently, based on a semiparametric dynamic latent space model, \cite{he2025semiparametric} established the uniform convergence rate for the estimated latent positions by solving an efficient score estimation equation. 



Although multilayer networks are frequently encountered in practice and theoretical results have been established for various methods, statistical inference for multilayer network models is still at its early stage, which is important in quantifying the estimation uncertainty for latent positions and layer-specific structures, as well as downstream tasks such as link prediction and network testing. Recently, a few works have established the inferential theory under the multilayer stochastic block model. In particular, \cite{arroyo2021inference} and \cite{xie2024bias} derived the asymptotic distributions for the eigenvectors and the layer-specific connection matrices under the common subspace independent edge (COSIE) random graph model, and \cite{su2026limit} obtained the asymptotic distributions for the connection matrices under the multilayer stochastic block model. However, both models assume that the binary edges follow linear models, which may not be the most suitable choice for more general settings where discrete edges are modeled through nonlinear link functions, as commonly encountered in many applications.




To see the challenge of the statistical inference for multilayer network models with nonlinear link functions, we note that the distribution of a multilayer network is determined by an underlying order-3 tensor with two node-wise dimensions and one layer-wise dimension, which, through nonlinear link functions, admits a low-rank Tucker decomposition under the multilayer latent space model discussed in Section~\ref{section:problem_setup}.
Then, the latent positions correspond to the loading matrices while the layer-specific connection matrices form the core tensor in the Tucker decomposition. 
Existing tensor inference has mainly focused on loading matrices under linear tensor models
\citep{xia2022inference, agterberg2024statistical}. Recent work on generalized
matrix factor models has further established joint-likelihood inference
for row and column loadings and time-specific factor matrices \citep{KongZhang2026}. 
Their formulation, however, does not incorporate layer-varying sender and
receiver degree heterogeneity, which is an important feature of
multilayer network data.

In this paper, we propose a generalized latent space model for directed multilayer
networks with continuous, binary, or count-valued edges through general
edge-specific distributions. The model combines sender and receiver
latent positions with layer-specific connection matrices and
layer-varying in- and out-degree effects, allowing it to capture both
interaction structures and heterogeneous node activity across layers. The edge distributions may vary across node pairs and layers, allowing heterogeneous types of relationships to be represented within a general framework.
It includes several existing multilayer network models as special cases \citep{zhang2020flexible, arroyo2021inference, he2025semiparametric, zhang2025efficient}.

For estimation, we develop a novel {\it Unfolding and Fusion} method that avoids direct solution of a highly nonconvex, large-scale tensor optimization problem. We aggregate the network layers into a tensor and consider its two node-mode unfoldings. For each unfolding, we estimate the underlying low-rank natural-parameter matrix through maximum likelihood. A two-sided centering operation then separates the structural latent spaces from the degree-effect subspaces, allowing the sender and receiver latent positions to be recovered from the two complementary unfoldings. The layer-specific connection matrices are subsequently reconstructed by fusing the corresponding estimated components. The resulting procedure reduces estimation to two low-rank matrix likelihood problems followed by matrix operations and permits consistent spectral initialization across the edge distributions considered.

For the latent positions, we establish consistency and asymptotic normality through a Lagrangian likelihood analysis of the unfolded low-rank models. Existing multilayer latent-space theory has primarily focused on consistency and convergence rates, whereas our results provide feasible node-specific inference for distinct sender and receiver latent positions in the presence of layer-varying degree heterogeneity. Because the factorization is invariant under transformations of the latent factors, the corresponding likelihood Hessian is singular and standard M-estimation arguments cannot be applied directly. To address this issue, we incorporate the identifiability constraints into a Lagrangian-adjusted likelihood and analyze the resulting Hessian to derive feasible limiting distributions. These results enable the construction of confidence regions for individual sender and receiver latent positions and support related downstream inferential tasks.

For the layer-specific connection matrices, the fusion construction gives rise to a new inferential challenge. Although each estimated component admits a first-order asymptotic expansion, the connection-matrix estimator is obtained by combining two high-dimensional estimated components. Their second-order estimation errors accumulate through the fusion operation and induce a non-negligible asymptotic bias on the same scale as the leading stochastic fluctuation. We derive an explicit representation of this bias and construct a feasible plug-in correction. After debiasing, the connection-matrix estimator is asymptotically normal, enabling valid confidence intervals and tests of equality between layer structures. More broadly, this phenomenon illustrates how higher-order errors that are individually negligible can become first-order relevant when high-dimensional factor estimators are combined to estimate low-dimensional parameters.

The remainder of the paper is organized as follows. Section~\ref{subsection:notations_and_preliminaries} introduces the notation and preliminary results on matrix and tensor operations. Section~\ref{section:problem_setup} describes the proposed model. Estimation procedures are developed in Section~\ref{section:estimation_and_inference}, and the corresponding theoretical results are presented in Section~\ref{section:theory}. Section~\ref{section:numerical_experiments} reports numerical results for both synthetic and real data. Section~\ref{section:conclusion} summarizes the main results and lists some future directions. Technical proofs and more numerical results are included in the Supplementary Material.


\subsection{Notation and preliminaries.}\label{subsection:notations_and_preliminaries}

We introduce some notation and definitions here, which are necessary throughout this paper and the supplement. For any integer $K$, let $[K]=\{1,\cdots,K\}$. We use $\operatorname{vec}(\cdot)$ to denote the column-wise vectorization of matrices, and $\otimes$ for the Kronecker product. The $\psi_1$-Orlicz norm of a random variable $\XX$, where $\psi_1(x)=\exp(x)-1$, is defined as 
$\|\XX\|_{\psi_1}:=\inf\{t>0:\EEE[\exp(|\XX|/t)]\le 2\}$. 
Note that $\XX$ is sub-exponential if $\|\XX\|_{\psi_1}<\infty$.
Let $\ee_q^{(d)}$ denote the $q$-th canonical basis vector in
$\mathbb{R}^d$, with the superscript omitted when the ambient dimension is clear.


For any $mr \times 1$ vector $\vv$ consisting of $m$ blocks of $r\times 1$ small vectors, $\left[\vv\right]_i = \vv_{(i-1)r+1:ir,1}$ for any $1\leq i \leq m$, where 
	$\left[\vv\right]_i$ represents the $i$-th $(r\times 1)$ sub-block of $\vv$. Moreover, for any $1\leq j\leq r,\,\left[ \left[\vv \right]_i \right]_j = \vv_{ (i-1)\cdot r + j,1}$. In a word, $\left[ \left[\vv \right]_i \right]_j$ denotes the $j$-th element of the $i$-th $(r\times 1)$ sub-block. 
For any matrix $\A$, let $\A_{i,j}$ be its $(i,j)$-th entry. We also denote $\A_{i,:}$ as its $i$-th row and $\A_{:,j}$ as the $j$-th column. The $\ell_{2,\infty}$-norm of matrix $\A$ is defined as $
    \|\A\|_{2,\infty}=\max_{i} \|\A_{i,:}\|_2$.

For any tensor $\XXX\in \mathbb{R}^{d_1\times d_2\times d_3}$, let $\XXX_{i,j,t}$ denote its $(i,j,t)$-th entry. Let $\mathcal{M}_m(\cdot)$ denote the mode-$m$ unfolding of a 3-dimensional tensor into a matrix. Specifically, $\mathcal{M}_1(\mathcal{X}) \in \mathbb{R}^{d_1 \times (d_2 d_3)}$ with $\XXX_{i_1,i_2,i_3} = [\MMM_1(\XXX)]_{i_1,i_2 + d_2(i_3-1)}$, and $\mathcal{M}_2(\mathcal{X}) \in \mathbb{R}^{d_2 \times (d_1 d_3)}$ with $\XXX_{i_1,i_2,i_3} = [\MMM_2(\XXX)]_{i_2,i_1 + d_1(i_3-1)}$.
For $n, T \in \mathbb{N}^{+}$, let $\AAA \in \mathbb{R}^{1 \times n \times T}$ be a tensor with $\MMM_2(\AAA)=\aaa \in \mathbb{R}^{n\times T}$, and $\BBB \in \mathbb{R}^{n \times 1 \times T}$ with $\MMM_1(\BBB)=\bbb \in \mathbb{R}^{n\times T}$. Then, the following unfolding identities hold:
\begin{equation}\label{eq:unfolding_intercepts}
\begin{aligned}
	\mathcal{M}_1(\AAA \times_1 \mathbf{1}_n) &= \mathcal{M}_1([\AAA; \mathbf{1}_n, \II_n, \II_T]) = \mathbf{1}_n \operatorname{vec}(\boldsymbol{\alpha})^\top, \\
	\mathcal{M}_2(\AAA \times_1 \mathbf{1}_n) &= \mathcal{M}_2([\AAA; \mathbf{1}_n, \II_n, \II_T]) = \boldsymbol{\alpha} (\II_T \otimes \mathbf{1}_n^\top), \\
	\mathcal{M}_3(\AAA \times_1 \mathbf{1}_n) &= \mathcal{M}_3([\AAA; \mathbf{1}_n, \II_n, \II_T]) = \boldsymbol{\alpha}^\top (\II_n \otimes \mathbf{1}_n^\top),\\
	\mathcal{M}_1(\BBB \times_2 \mathbf{1}_n) &= \mathcal{M}_1([\BBB; \II_n, \mathbf{1}_n, \II_T]) = \boldsymbol{\beta} (\II_T \otimes \mathbf{1}_n^\top), \\
	\mathcal{M}_2(\BBB \times_2 \mathbf{1}_n) &= \mathcal{M}_2([\BBB; \II_n, \mathbf{1}_n, \II_T]) = \mathbf{1}_n \operatorname{vec}(\boldsymbol{\beta})^\top.\\
\end{aligned}
\end{equation}

\section{Proposed Method}\label{section:problem_setup}

\subsection{Generalized
Multilayer Latent Space Model}
Let $\mathcal G$ denote a multilayer directed network comprising $T$ network layers on $n$ common nodes. For any $t\in [T]$, the $t$-th network layer can be represented via its adjacency matrix $(y_{ijt})_{n\times n}$,
where $y_{ijt}$ indicates the directed interaction from node $i$ to node $j$ on the $t$-th layer.
We assume each $y_{ijt}$ is generated from 
\begin{equation}\label{eq:model}
y_{ijt} \sim g_{ijt}(\cdot \mid x_{ijt}),~~\text{with}~~ x_{ijt} = \ttt_i^\top\LLLL_t\ph_j + \beta_{it} + \alpha_{jt},
\end{equation} 
where $g_{ijt}(\cdot\mid\cdot)$ is some known probability density/mass function, which is allowed to vary across different node pairs $(i,j)$ and layers $t$. Here the directed interaction effect from node $i$ to node $j$ on the $t$-th layer is modeled through $\ttt_i^\top\LLLL_t\ph_j$, where $\ttt_i\in\R^{k_1}$ and $\ph_j\in\R^{k_2}$ are respectively the latent positions capturing the sending and receiving behavior of nodes $i$ and $j$, and $\LLLL_t \in \mathbb{R}^{k_1\times k_2}$ is a layer-specific connection matrix governing the interaction structure of the $t$-th layer. Further, $\beta_{it}$ and $\alpha_{jt}$ are respectively the out-degree and in-degree heterogeneity parameters for nodes $i$ and $j$ on the $t$-th layer. We suppose all $y_{ijt}$ are independent conditioning on $\{\ttt_i\}_{i=1}^n,~\{\ph_j\}_{j=1}^n,~\{\alpha_{it},\beta_{it}\}_{i\in[n],t\in[T]}$ and $\{\LLLL_t\}_{t=1}^T$.


\begin{remark}\label{rk:types}
    The flexibility on the distribution functions $g_{ijt}$'s allows us to deal with networks with different kinds of edge types, including binary, count, and continuous data types. For such types of edges, the exponential family of distributions can be used for modeling \citep{rabe2004generalized}, whose density takes the form $g_{ijt}(y) = \exp\{(yx_{ijt} - b_{ijt}(x_{ijt}))/\phi_{ijt} - c_{ijt}(y,\phi_{ijt})\}$, where $b_{ijt},c_{ijt}$ are known functions and $\phi_{ijt}$ is the dispersion parameter. If $y_{ijt}$ is continuous, we may assume $g_{ijt}$ is a Gaussian density function, where $\phi_{ijt}$ is the variance, $b_{ijt}(x_{ijt}) = x_{ijt}^2/2$ and $c_{ijt}(y_{ijt}, \phi_{ijt}) = y_{ijt}^2/(2\phi_{ijt}) + \frac{1}{2}\log(2\pi\phi_{ijt})$. For binary $y_{ijt}\in\{0,1\}$, we can assume it follows a logistic model with $\phi_{ijt} = 1,~b_{ijt}(x_{ijt}) = \log\bigl(1 + e^{x_{ijt}}\bigr)$ and $c_{ijt}(y_{ijt}, \phi_{ijt}) = 0$. A Poisson model may be assumed if $y_{ijt}$ is a count variable, where $\phi_{ijt} = 1$, $b_{ijt}(x_{ijt}) = e^{x_{ijt}},~c_{ijt}(y_{ijt}, \phi_{ijt}) = \log(y_{ijt}!)$.
\end{remark}

\begin{remark}\label{special_models}
The model \eqref{eq:model} includes many existing network models as special cases. 
For undirected networks, we set $k_1=k_2, \ttt_i = \ph_i$ and $\beta_{it} = \alpha_{it}$. If all $y_{ijt}$ for $i\neq j$ are Bernoulli random variables with mean $1/(1+e^{-x_{ijt}})$, then \eqref{eq:model} becomes the multilayer latent space model in \cite{zhang2020flexible}. 
If all $y_{ijt}$ are Bernoulli random variables with mean $x_{ijt}$ and $\beta_{it} = \alpha_{it} = 0$, then \eqref{eq:model} reduces to the COSIE random graph model \citep{arroyo2021inference}.
If all $y_{ijt}$ are Poisson random variables with mean $e^{x_{ijt}}$ and $\LLLL_t = \II_k$, then \eqref{eq:model} is equivalent to the Poisson-based latent space model in \cite{he2025semiparametric}. 
For directed networks, if all $y_{ijt}$ are Poisson random variables with mean $e^{x_{ijt}}$ with $\beta_{it} = \alpha_{jt} \equiv \text{const}$ and $\text{vec}(\LLLL_t) = \MM\ww_t$ for some $\MM\in\R^{k_1k_2\times r}$ and $\ww_t\in\R^{r}$, then \eqref{eq:model} is the same as the dynamic network embedding model in \cite{zhang2025efficient}.
\end{remark}

Let $\XXX = (x_{ijt})_{n\times n\times T}$, $\TTT = (\ttt_1,...,\ttt_n)^\top$ and $\PH = (\ph_1,...,\ph_n)^\top$. For $t\in[T]$, define $\XX_t = (x_{ijt})_{n\times n}$, $\aaa_t = (\alpha_{1t},...,\alpha_{nt})^\top, \bbb_{t} = (\beta_{1t},...,\beta_{nt})^\top$. Further define $\aaa = (\aaa_1,...,\aaa_T)$ and $\bbb = (\bbb_1,...,\bbb_T)$. Then, 
\begin{equation}\label{eq:basic_model_setup}
\XX_t = \TTT \LLLL_t \PH^\top + \mathbf{1}_n \aaa_t^\top + \bbb_t \mathbf{1}_n^\top,\quad t\in[T].
\end{equation} 
To facilitate estimation, it is helpful to represent $\XXX$ in a tensor product form. 
To this end, define tensors $\AAA \in \mathbb{R}^{1 \times n \times T}$ and $\BBB \in \mathbb{R}^{n \times 1 \times T}$ such that $\AAA_{1,:,:} = \aaa$ and $\BBB_{:,1,:} = \bbb$.
Further, define a tensor $\mathcal{S} \in \mathbb{R}^{k_1 \times k_2 \times T}$ such that $\SSSS_{:,:,t} = \LLLL_t$ for $t\in[T]$. Then, by \eqref{eq:unfolding_intercepts} and \eqref{eq:basic_model_setup},
\begin{equation}\label{eq:tensor-decomp}
\XXX = [\mathcal{S}; \boldsymbol{\Theta}, \PH, \II_T] + [\AAA; \mathbf{1}_n, \II_n, \II_T] + [\BBB; \II_n, \mathbf{1}_n, \II_T].
\end{equation}

To model degree heterogeneity in a parsimonious way, we impose a fixed-rank factor structure on
$\aaa\in\R^{n\times T}$ and $\bbb\in\R^{n\times T}$.
Specifically, we assume that there exist
$\UU_\alpha\in\R^{n\times k_\alpha}$,
$\UU_\beta\in\R^{n\times k_\beta}$,
$\VV_\alpha\in\R^{T\times k_\alpha}$, and
$\VV_\beta\in\R^{T\times k_\beta}$, with fixed
$k_\alpha$ and $k_\beta$, such that
\begin{equation}\label{eq:degree decomp}
    \aaa=\UU_\alpha\VV_\alpha^\top
    \quad\text{and}\quad
    \bbb=\UU_\beta\VV_\beta^\top.
\end{equation}
This specification allows heterogeneous node-specific loadings on a
small number of common cross-layer degree profiles, such as global density shifts, temporal shocks, or persistent sender/receiver activity patterns. 

\begin{remark}
The factor restriction in \eqref{eq:degree decomp} is introduced to
simplify estimation and is not essential to the general formulation of
degree heterogeneity. Unrestricted node- and layer-specific degree effects
can alternatively be handled through profiling or semiparametric
adjustment. For example, \citet{he2025semiparametric} consider a
longitudinal Poisson latent space model with unrestricted time-varying
node-specific baseline effects and develop estimation procedures for a
shared, time-invariant latent interaction structure. Their model,
however, does not contain layer-specific connection matrices, and their
theoretical results focus on estimation error bounds rather than
statistical inference for layer-specific interaction structures.
Our primary objective is statistical inference for the latent positions
and, in particular, for the layer-specific connection matrices.
We therefore adopt \eqref{eq:degree decomp} as a tractable representation
of the degree effects and focus on the inferential challenges
arising from the layer-varying interaction structure. The resulting
estimation procedure is developed in
Section~\ref{section:estimation_and_inference}, and the corresponding
debiased inference for the connection matrices is presented in
Section~\ref{subsection:estimation_asymp_dist_core}.
Similar factor structures are standard in large panel models when both $n$ and $T$ are large \citep{bai2012statistical, fan2016projected, chen2020structured}. 
Empirical evidence supporting this modeling choice is provided in Section~\ref{subsection:real_data}. 
\end{remark}


According to \eqref{eq:unfolding_intercepts} and \eqref{eq:tensor-decomp}, unfolding $\XXX$ along the first mode yields 
\begin{equation}\label{eq:M1X_sum}
\begin{aligned}
	\mathcal{M}_1(\mathcal{X}) &= \boldsymbol{\Theta} \mathcal{M}_1(\mathcal{S}) \left(\II_T \otimes \PH^\top \right) + \boldsymbol{\beta} (\II_T \otimes \mathbf{1}_n^\top) + \mathbf{1}_n 	\operatorname{vec}(\aaa)^\top \\
	&= \begin{pmatrix} \boldsymbol{\Theta} & \UU_\beta & \mathbf{1}_n \end{pmatrix}
	   \begin{pmatrix}
	   	\mathcal{M}_1(\mathcal{S}) \left(\II_T \otimes \PH^\top\right) \\
	   	\VV_\beta^\top (\II_T \otimes \mathbf{1}_n^\top) \\
	   	\operatorname{vec}(\aaa)^\top
	   \end{pmatrix} =: \UU_1 \VV_1^\top,
\end{aligned}
\end{equation}
with 
\begin{equation}\label{eq:def_U_1_V_1}
\UU_1 := \begin{pmatrix} \boldsymbol{\Theta} & \UU_\beta & \mathbf{1}_n \end{pmatrix},\quad \VV_1:= 
\begin{pmatrix}
	(\II_T\otimes \PH)\MMM_1(\SSSS)^\top & (\II_T\otimes \mathbf{1}_n)\VV_\beta & \operatorname{vec}(\aaa)
\end{pmatrix},
\end{equation}
where $\UU_1 \in \mathbb{R}^{n \times d_1}$ and $\VV_1 \in \mathbb{R}^{(nT) \times d_1}$ with $d_1 :=k_1 + k_\beta + 1$ constitute a low-rank factorization of $\MMM_1(\XXX)$. Similarly, unfolding $\mathcal{X}$ along the second mode gives
\begin{equation}\label{eq:M2X_sum}
\begin{aligned}
	\mathcal{M}_2(\mathcal{X}) & = \PH \mathcal{M}_2(\mathcal{S}) \left(\II_T \otimes \boldsymbol{\Theta}^\top \right) + \boldsymbol{\alpha} (\II_T \otimes \mathbf{1}_n^\top) + \mathbf{1}_n \operatorname{vec}(\boldsymbol{\beta})^\top\\
	&= \begin{pmatrix} \PH & \UU_\alpha  & \mathbf{1}_n\end{pmatrix}
	   \begin{pmatrix}
	   	\mathcal{M}_2(\mathcal{S}) (\II_T \otimes \boldsymbol{\Theta}^\top) \\
	   	\VV_\alpha^\top (\II_T \otimes \mathbf{1}_n^\top)\\
        	\operatorname{vec}(\boldsymbol{\beta})^\top \\
	   \end{pmatrix} =: \UU_2 \VV_2^\top,
\end{aligned}
\end{equation}
where $\UU_2\in \mathbb{R}^{n \times d_2}$ and $\VV_2\in \mathbb{R}^{nT \times d_2}$ with $d_2 :=k_2 + k_\alpha + 1$ represent the corresponding low-rank factors in $\MMM_2(\XXX)$, with explicit forms:
\begin{equation}\label{eq:def_U_2_V_2}
\UU_2 := 
\begin{pmatrix}
    \PH& \UU_\alpha & \mathbf{1}_n 
\end{pmatrix}, \quad
\VV_2 :=
\begin{pmatrix}
    (\II_T \otimes \boldsymbol{\Theta}) \mathcal{M}_2(\mathcal{S})^\top&
    (\II_T \otimes \mathbf{1}_n) \VV_\alpha &
    \operatorname{vec}(\boldsymbol{\beta}) 
\end{pmatrix}.
\end{equation}
%


\subsection{Identifiability}\label{subsection:identifiability}

Note that there is an identifiability issue for the model parameters. 
Let $\TTT^*,~\PH^*,~\UU_{\alpha}^*,~\UU_{\beta}^*$, $\VV_{\alpha}^*,~\VV_{\beta}^*,~\SSSS^*$ and $\XXX^*$ be the true parameter matrices and tensors from \eqref{eq:tensor-decomp} and \eqref{eq:degree decomp}. 
Then, for any invertible matrices $\FF_1\in\R^{k_1\times k_1}$ and $\FF_2\in\R^{k_2\times k_2}$, we have $$
[\SSSS^*;\TTT^*,\PH^*,\II_T] = [\SSSS^*\times_1 \FF_1^{-1}\times_2 \FF_2^{-1};\TTT^*\FF_1,\PH^*\FF_2,\II_T],
$$ which makes it impossible to distinguish $(\SSSS^*,\TTT^*,\PH^*)$ and $(\SSSS^*\times_1 \FF_1^{-1}\times_2 \FF_2^{-1},\TTT^*\FF_1,\PH^*\FF_2)$ without further conditions.
We now establish the identifiability conditions for the parameters of interest, specifically the $\TTT^*,\PH^*$ and $\SSSS^*$.
Let us first consider $\TTT^*$ and $\PH^*$. 
 Define the centering matrix $\mathbf{J}_n = \II_n - n^{-1}\mathbf{1}_n\mathbf{1}_n^\top$ and the block-wise centering matrix $\mathbf{J}_{n,T}=\II_T\otimes \mathbf{J}_n$. We start by noting that, from \eqref{eq:def_U_1_V_1} and \eqref{eq:def_U_2_V_2}, 
\begin{equation}\label{eq:centering}
\JJ_n\MMM_1(\XXX^*)\JJ_{n,T}^\top = \TTT^*\MMM_1(\SSSS^*)(\II_T\otimes \PH^{*\top}),\quad \JJ_n\MMM_2(\XXX^*)\JJ_{n,T}^\top = \PH^*\MMM_2(\SSSS^*)(\II_T\otimes \TTT^{*\top}),    
\end{equation}
which implies that $\TTT^*$ and $\PH^*$ could be identified from the column spaces of $\JJ_n\MMM_1(\XXX^*)\JJ_{n,T}^\top$ and $\JJ_n\MMM_2(\XXX^*)\JJ_{n,T}^\top$, respectively, under some conditions. We now specify such conditions in the following Assumption~\ref{assumption:identi}. For $m\in[2]$, let $\UU_m^*$ and $\VV_m^*$ be the true values of $\UU_m$ and $\VV_m$ from \eqref{eq:def_U_1_V_1} and \eqref{eq:def_U_2_V_2}. 


    




\begin{assumption}\label{assumption:identi}
For $m\in [2]$, let $\UU_m^*\VV_m^{*\top} = \UU_m^{svd}\VV_m^{svd\top}$ be a decomposition such that $\UU_m^{svd\top}\UU_m^{svd} = n \II_{d_m},\VV_m^{svd\top}\VV_m^{svd}/nT$ is a diagonal matrix with decreasing entries, where $\UU_m^{svd}\in \mathbb{R}^{n\times d_m}$ and $\VV_m^{svd}\in \mathbb{R}^{nT\times d_m}$. 
\begin{enumerate}
    \item \label{assumption:Theta_Psi_center_intercepts_orthogonal}
    $\boldsymbol{\Theta}^{*\top} \boldsymbol{\Theta}^* = n\II_{k_1}$ and $\PH^{*\top} \PH^* = n\II_{k_2}$; $\TTT^*\perp \operatorname{span}\{\mathbf{1}_n, \UU_\beta^*\}$ and $\PH^*\perp \operatorname{span}\{ \mathbf{1}_n, \UU_\alpha^*\}$. 

    
	\item \label{assumption:distinct_eigenvalues}
	For $m\in[2]$, $\MMM_m(\SSSS^*)\MMM_m(\SSSS^*)^\top/T$ are diagonal matrices with strictly decreasing positive diagonals, and so do the limits as $T\to\infty$.

    \item For $m\in[2]$, the diagonal entries of $\VV_m^{svd\top}\VV_m^{svd}/nT$ are positive and different, and so do the limits as $nT\to\infty$.

    \item For $m\in[2]$, $\|\UU_m^{svd}\|_{2\to\infty}\leq C$ and $\|\VV_m^{svd}\|_{2\to\infty}\leq C$ for a constant $C$.
\end{enumerate}
\end{assumption}

\begin{remark}
Conditions in Assumption~\ref{assumption:identi} are standard and consistent with the literature of network analysis and tensor estimation \citep{xia2022inference, li2023statistical, agterberg2024statistical, he2025semiparametric}.
Note that $\UU_m^{svd}$ and $\VV_m^{svd}$ are the left and right singular matrices for $\UU_m^*\VV_m^{*\top}$ respectively after column scaling. The first condition in Assumption~\ref{assumption:identi} requires that the columns of $\TTT^*$ and $\PH^*$ are orthogonal and centered, and also orthogonal to column spaces of the intercept matrices, $\aaa^*$ and $\bbb^*$, respectively. This is a crucial condition to identify $\TTT^*$ and $\PH^*$ from $\UU_m^{svd}$ for $m\in[2]$.
The second condition aims to fix the rotations after the column spaces of $\JJ_n\MMM_1(\XXX)\JJ_{n,T}^\top$ and $\JJ_n\MMM_2(\XXX)\JJ_{n,T}^\top$ are obtained. 
The third condition requires eigengaps between consecutive singular values of $\UU_m^*\VV_m^{*\top}$, which is a common assumption in the literature on matrix perturbation \citep{yu2015useful}. The last condition in Assumption~\ref{assumption:identi} assumes that each row of $\UU_m^{svd}$ and $\VV_m^{svd}$ are constrained in a compact set, which is also common and relates to the incoherence condition in matrix completion \citep{candes2012exact}.
\end{remark}


Under Assumption~\ref{assumption:identi}, the columns of $\TTT^*$ and $\PH^*$ span the left singular subspaces of $\MMM_1(\XXX^*)$ and $\MMM_2(\XXX^*)$, respectively. We formalize this in Lemma~\ref{lemma:index_set}. 

\begin{lemma}\label{lemma:index_set}
    For $m\in[2]$, there exist ordered index subsets $S_m^*\subseteq [d_m]$ with $|S_m^*| = k_m$ and diagonal sign matrices $\RR_m$ with entries in $\{\pm 1\}$ such that \[
    \begin{aligned}
\TTT^* =& [\UU_1^{svd}\RR_1]_{:,S_1^*},\quad (\II_T\otimes \boldsymbol{\Phi}^*)\MMM_1(\SSSS^*)^\top = [\VV_1^{svd}\RR_1]_{:,S_1^*}, \\ 
\PH^* =& [\UU_2^{svd}\RR_2]_{:,S_2^*},\quad (\II_T\otimes \boldsymbol{\Theta}^*)\MMM_2(\SSSS^*)^\top = [\VV_2^{svd}\RR_2]_{:,S_2^*}.
\end{aligned}
\] 
\end{lemma}

We now turn to $\SSSS^*$. After recovering $\TTT^*$ and $\PH^*$, the $\MMM_1(\SSSS^*)(\II_T\otimes \PH^{*\top})$ and $\MMM_2(\SSSS^*)(\II_T\otimes \TTT^{*\top})$ are also obtained, denoted by $(\VV_1^{c*})^\top$ and $(\VV_2^{c*})^\top$. The identifiability of $\SSSS^*$ is implied from the following relationship 
\begin{equation}\label{eq:core_form}
  (\VV_1^{c*})^\top (\II_T\otimes \PH^*) = \MMM_1(\SSSS^*)(\II_T\otimes \PH^{*\top})(\II_T\otimes \PH^*) = \MMM_1(\SSSS^*)(\II_T \otimes (\PH^{*\top}\PH^*)) = n\MMM_1(\SSSS^*),  
\end{equation}
where the last equality is due to condition 1 in Assumption~\ref{assumption:identi}. We formalize the identifiability result in the following proposition.

\begin{proposition}\label{proposition:identi}
    Suppose Assumption~\ref{assumption:identi} holds. For any set of parameters $(\check\XXX, \check\TTT,\check\PH,\check\SSSS,\check\aaa,\check\bbb)$ that satisfies \eqref{eq:tensor-decomp}, \eqref{eq:degree decomp} and Assumption~\ref{assumption:identi}, if $\check\XXX = \XXX^*$, then $\check\TTT = \TTT^*\RR_1,~\check\PH = \PH^*\RR_2$ and $\check\SSSS = [\SSSS^*;\RR_1,\RR_2,\II_T]$, where $\RR_1\in\R^{k_1\times k_1}$ and $\RR_2\in\R^{k_2\times k_2}$ are diagonal matrices with diagonal entries being $\pm1$. 
\end{proposition}

\begin{remark}\label{rmk:identifiability_and_signed}
We give a discussion on the result of Proposition~\ref{proposition:identi}. First, we could only expect to identify $\TTT^*$ and $\PH^*$ up to some column sign flipping, represented by $\RR_1$ and $\RR_2$, which also appear in identifying $\SSSS^*$. However, this sign uncertainty will not affect the inference task such as testing if $[\SSSS^*]_{i,j,t} - [\SSSS^*]_{i,j,t^\prime} = 0$, for $i\in[k_1]~,j\in[k_2]$ and $t\neq t^\prime$. This is because $\RR_1$ and $\RR_2$ affect $\SSSS^*$ only through the first two modes. Detailed result is given in Section~\ref{section:theory}. Second, the challenging part in the current formulation \eqref{eq:tensor-decomp} is the identification and estimation for the tensor components $\TTT^*,\PH^*$ and $\SSSS^*$, which is new in the literature and that is what we will focus on in the following.
The intercept parts, $\boldsymbol{\alpha}^*$ and $\boldsymbol{\beta}^*$, are more similar to previous factor analysis \citep{chen2020structured, wangMaximumLikelihoodEstimation2022}, so we here only focus on the new and more challenging parts of $\TTT^*,\PH^*$ and $\SSSS^*$.
\end{remark}


\subsection{Estimation}\label{section:estimation_and_inference}

According to \eqref{eq:tensor-decomp}, we need to estimate the loading matrices and core tensor from a tensor Tucker decomposition based on $\{y_{ijt}\}_{i,j\in[n],t\in[T]}$, which is a challenging task since a large-scale tensor optimization is highly non-convex and computationally intensive, especially when $y_{ijt}$'s are of different types. In the following, we propose a novel {unfolding and fusion} estimation procedure.

Let $\Z_m \in \mathbb{R}^{n \times nT}$ denote the mode-$m$ unfolding of $\XXX$ for $m\in[2]$. From \eqref{eq:def_U_1_V_1} and \eqref{eq:def_U_2_V_2}, $\Z_m = \UU_m \VV_m^\top$. For $i\in[n]$ and $j\in[nT]$, let $\ell_{m,i,j}([\Z_m]_{i,j})$ denote the log-likelihood based on the observation $[\MMM_m(\YYY)]_{i,j}$ with $\YYY = (y_{ijt})_{n\times n\times T}$ being the adjacency tensor.
Then, for $m\in [2]$, the log-likelihood function corresponding to mode-$m$ unfolding takes the form
\begin{equation}\label{eq:log_likelihood_m_mode}
    L_m(\Z_m) = \sum_{i \in [n]} \sum_{j \in [nT]} \ell_{m,i,j}([\Z_m]_{i,j}).
\end{equation}
We estimate $(\UU_m, \VV_m)$ by solving the following constrained maximum likelihood problem:
\begin{equation} \label{op:original_OP}
\begin{aligned}
(\widehat{\UU}_m, \widehat{\VV}_m) =& \argmax_{\UU_m,\, \VV_m}   L_m(\UU_m \VV_m^\top) \\
\text{subject to} \quad
& \|\UU_m\|_{2 \to \infty} \leq C,\,\|\VV_m\|_{2 \to \infty} \leq C,\\
& \UU_m^\top \UU_m = n\II_{d_m}, \\
& \VV_m^\top \VV_m \text{ is diagonal with decreasing entries}.
\end{aligned}
\end{equation}


\begin{remark}\label{rk:likelihood_identifi}
Note that the log-likelihood function $L_m$ depends on $(\UU_m,\VV_m)$ only through $\UU_m\VV_m^\top$. Therefore, we could rotate $\UU_m$ and $\VV_m$ simultaneously by multiplying an orthogonal matrix without changing the value of $L_m$. The last two constraints in \eqref{op:original_OP} are to fix the rotation. We also require each row of $\UU_m$ and $\VV_m$ lies within a compact set. Similar constraints have been adopted in the literature of factor analysis \citep{bai2003inferential, wangMaximumLikelihoodEstimation2022}.
\end{remark}

Based on \eqref{eq:centering}, we further apply two-sided centering to $\widehat{\Z}_m = \widehat\UU_m\widehat\VV_m^\top$ to get $\JJ_n\widehat{\Z}_m\JJ_{n,T}^\top$. Then, we identify the columns of $\widehat\UU_1$ and $\widehat\UU_2$ which correspond to $\TTT$ and $\PH$ respectively by choosing the $k_m$ largest norm of columns after projection onto the column space of $\JJ_n\widehat{\Z}_m\JJ_{n,T}^\top$. The $\widehat\TTT$ and $\widehat\PH$ are obtained by selecting such columns from $\widehat\UU_1$ and $\widehat\UU_2$, respectively. Finally, the $\{\LLLL_t\}_{t=1}^T$ is estimated according to \eqref{eq:core_form}. We summarize the estimation procedure in Algorithm~\ref{alg:structure_recovery} below.

\begin{algorithm}[H]
\caption{Estimating latent positions and connection matrices by {\it unfolding and fusion}.}\label{alg:structure_recovery}
\begin{algorithmic}[1]
\Require Adjacency tensor $\YYY = (y_{ijt})_{n\times n\times T}$, link functions $\{g_{ijt}(\cdot):i,j\in[n],t\in[T]\}$,
dimensions $d_1,d_2,k_1,k_2$.
\For{$m=1,2$}
  \State Obtain $(\widehat{\UU}_m,\widehat{\VV}_m)$ from the constrained maximum likelihood optimization \eqref{op:original_OP} based on mode-$m$ unfolding.
  \State Form $\widehat{\Z}_m=\widehat{\UU}_m\widehat{\VV}_m^\top$ and its centered version $\widehat{\Z}_m^c=\JJ_n\widehat{\Z}_m\JJ_{n,T}^\top$. Define $\PP_m\in\R^{n\times n}$ as the projection matrix onto the column space of $\widehat{\Z}_m^c$.

  

  \State For each $j\in [d_m]$, compute the projection norm $s_{m,j}=\|[\PP_m\,\widehat{\UU}_m]_{:,j}\|_2$. 
  Select $\widehat{S}_m$ as the ordered index set corresponding to the $k_m$ largest values among $\{s_{m,j}\}$, and arrange the elements of $\widehat{S}_m$ such that the diagonals of $[\widehat{\VV}_m]_{:,\widehat{S}_m}^\top[\widehat{\VV}_m]_{:,\widehat{S}_m}$ are in decreasing order.
  \State Set $\widehat{\UU}_m^c=[\widehat{\UU}_m]_{:,\widehat{S}_m}$, $\widehat{\VV}_m^c= [\widehat{\VV}_m]_{:,\widehat{S}_m}$, and $\widehat\SSSS\in\R^{k_1 \times k_2\times T}$ by $$
  \mathcal{M}_1(\widehat{\mathcal{S}})=(\widehat{\VV}_1^c)^\top (\II_T \otimes \widehat\UU_2^c)/n.
  $$
\EndFor
\State Set $\widehat \TTT = \widehat\UU_1^c,~\widehat\PH = \widehat\UU_2^c$ and $\{\widehat\LLLL_t\}_{t=1}^T$ with $\widehat\LLLL_t = \widehat\SSSS_{:,:,t}$.
\Ensure $\{\widehat \TTT, \widehat\PH, \{\widehat\LLLL_t\}_{t=1}^T,\{\widehat{\VV}_m^c\}_{m\in [2]},\{\widehat S_m\}_{m\in[2]}\}$
\end{algorithmic}
\end{algorithm}

To estimate the latent positions, Algorithm 1 solves two constrained maximum likelihood optimizations regarding low-rank matrices through tensor unfolding.
This enables us to utilize the rich literature of factor analysis and low-rank matrix estimation. Specifically, there exist consistent initial values by spectral methods for different data types \citep{bai2003inferential, zhang2020note, he2025semiparametric}. This is in sharp contrast to solving an optimization for a large-scale tensor, which is highly non-convex and hard to find consistent initial values, especially for non-continuous edge types \citep{han2022optimal}. Furthermore, by tensor unfolding, the dimensions $d_1,d_2$ could be consistently estimated \citep{zhang2020note, chen2022determining}. In practice, $k_1,k_2$ could be chosen by the scree plots for singular values of $\widehat\Z_m^c$ for $m\in[2]$.
Finally, Algorithm 1 estimates the connection matrices, which correspond to the core tensor in the Tucker decomposition, by a fusion operator which involves only matrix multiplication. 
The fusion step is computationally simple and avoids direct nonlinear
tensor optimization. Statistically, however, it combines two estimated
factor components and therefore produces a non-negligible second-order
bias. Section~\ref{subsection:estimation_asymp_dist_core} explicitly characterizes and corrects this bias.

\section{Theory}\label{section:theory}

In this section, we establish the convergence rates and asymptotic distributions for the latent position and the connection matrix estimators, which further facilitate statistical inference tasks for multilayer networks. 

\subsection{Asymptotics for the latent position estimators}

For $m\in [2],i\in[n]$ and $j\in[nT]$, denote $\pi_{m,i,j}^* = [\MMM_m(\XXX^*)]_{i,j}$. We impose the following regularity conditions on the likelihood functions. 

\begin{assumption}\label{assumption:regularities-on-ell}
For any given compact domain $\Pi$, there exists $M>0$ such that for all $m\in [2],~i\in[n]$, $j\in [nT]$: 1, there exist $0 < b_L < b_U$ such that $b_L \le -\frac{\partial^2 \ell_{m,i,j}(\pi)}{\partial \pi^2} \le b_U$ for all $\pi \in \Pi$; 2, $\sup_{\pi \in \Pi} \left| \frac{\partial^3\ell_{m,i,j}(\pi)}{\partial \pi^3}  \right| \le M$ and $\sup_{\pi \in \Pi} \left| \frac{\partial^4\ell_{m,i,j}(\pi)}{\partial \pi^4} \right| \le M$; 3, $\|\frac{\partial \ell_{m,i,j}(\pi_{m,i,j}^*)}{\partial \pi}\|_{\psi_1} \leq M$. 
\end{assumption}


\begin{remark}
    Assumption~\ref{assumption:regularities-on-ell} specifies smoothness conditions on the individual log-likelihood function $\ell_{m,i,j}(\cdot)$, which are commonly used in the literature \citep{bai2003inferential, wangMaximumLikelihoodEstimation2022}. The first and second conditions in Assumption~\ref{assumption:regularities-on-ell} ensure the concavity for each $\ell_{m,i,j}(\cdot)$ and the boundedness of its third and
fourth-order derivatives, when true parameters lie in a compact set. The third condition states that the score functions are sub-exponential, which is important in establishing the error bounds for the estimators. Assumption~\ref{assumption:regularities-on-ell} holds for many types of $g_{ijt}(\cdot\mid x_{ijt})$ in \eqref{eq:model}, including linear, logistic, probit and Poisson models. 
\end{remark}


We now study the asymptotic behavior of $\widehat{\UU}_m$ and $\widehat{\VV}_m$ for $m \in [2]$. 
The following Theorem~\ref{thm:loading_individual_consistency} establishes the uniform convergence for the estimated latent positions.

\begin{theorem}\label{thm:loading_individual_consistency}
Suppose Assumptions~\ref{assumption:identi}-\ref{assumption:regularities-on-ell} hold. Then, as $n\to \infty$, for any fixed sufficiently small $\delta>0$,
\begin{equation}\label{eq:F norm rate}
\|\widehat\TTT-\TTT^*\RR_1\|_F^2=
O_p\left(\frac{1}{T}+\frac{T^{2\delta}}{n^{1-4\delta}}\right),\quad 
 \|\widehat\PH-\PH^*\RR_2\|_F^2
=
O_p\left(\frac{1}{T}+\frac{T^{2\delta}}{n^{1-4\delta}}\right),
\end{equation}
where $\RR_m\in\mathbb{R}^{k_m\times k_m}$ are diagonal with entries in $\pm 1$.
Moreover, the uniform convergence rates for individualized estimation error are
{\footnotesize 
\[
\|\widehat\TTT-\TTT^*\RR_1\|_{2\to\infty}=
O_p\left(
\log n\cdot \left\{\frac{1}{\sqrt{nT}}+\frac{T^\delta}{n^{1-2\delta}}\right\}
\right),\quad 
 \|\widehat\PH-\PH^*\RR_2\|_{2\to\infty}
=
O_p\left(
\log n\cdot \left\{\frac{1}{\sqrt{nT}}+\frac{T^\delta}{n^{1-2\delta}}\right\}
\right),
\]}
\end{theorem}


\begin{remark}
The convergence rates for the latent positions estimators in Frobenius norm are consistent with \cite{arroyo2021inference} and \cite{he2025semiparametric}, both of which focus on undirected multilayer networks. \eqref{eq:F norm rate} is also consistent with the result in \cite{zhang2020flexible} in the special case when there are no degree heterogeneity parameters.
Theorem~\ref{thm:loading_individual_consistency} further establishes the uniform convergence rates for the latent positions estimation errors. When $T=1$, the rate becomes $O_p(n^{-1/2}\log n)$, and is slightly faster than the rate derived in \cite{li2023statistical} for single layer latent space model, which is $O_p(n^{-1/2+\eta})$ for arbitrarily small $\eta > 0$.
\end{remark}


We now focus on the asymptotic distributions of $[\widehat{\boldsymbol{\Theta}}]_{i,:}$ 
and $[\widehat{\boldsymbol{\Phi}}]_{i,:}$ for $i\in[n]$. 
Recall the definitions of $\UU_m^{svd},\VV_m^{svd}$ in Assumption~\ref{assumption:identi}. For $m \in [2]$, we further define $\uu_m^{svd} = \ve((\UU_m^{svd})^\top)$ and $\vv_m^{svd} = \ve((\VV_m^{svd})^{\top})$, such that 
\begin{equation}\label{eq:def_uu_m_vv_m}
\begin{aligned}
\uu_m^{svd} =& ([\uu_m^{svd}]_1^\top, \ldots, [\uu_m^{svd}]_n^\top)^\top \in \mathbb{R}^{nd_m}, \\
\vv_m^{svd} =& ([\vv_m^{svd}]_1^\top, \ldots, [\vv_m^{svd}]_{nT}^\top)^\top \in \mathbb{R}^{nTd_m},
\end{aligned}
\end{equation}
where $[\uu_m^{svd}]_s^\top,[\vv_m^{svd}]_s^\top\in \R^{1\times d_m}$ are the $s$-th rows of $\UU_m^{svd}$ and $\VV_m^{svd}$, respectively. 
For any $i\in [n],j\in [nT]$, denote 
\begin{equation}\label{eq:def_Sigma_Omega_u}
\small
\boldsymbol{\Sigma}_{\uu_m^{svd},j}:=  \sum_{s\in [n]} \left(-\frac{\partial^2\ell_{m,s,j}(\pi_{m,s,j}^*)}{\partial \pi_{m,s,j}^2}\right)[\uu_{m}^{svd}]_{s}[\uu_{m}^{svd}]_{s}^\top, \quad \boldsymbol{\Omega}_{\uu_m^{svd},j}:= \sum_{s\in [n]}\left(\frac{\partial \ell_{m,s,j}(\pi_{m,s,j}^*)}{\partial \pi_{m,s,j}}\right)^2[\uu_{m}^{svd}]_{s}[\uu_{m}^{svd}]_{s}^\top,
\end{equation}
\begin{equation}\label{eq:def_Sigma_Omega_v}
\small
{\boldsymbol{\Sigma}}_{{\vv}_m^{svd}, i} := \sum_{s\in[nT]}\left(-\frac{\partial^2 \ell_{m,i,s}(\pi_{m,i,s}^*)}{\partial \pi_{m,i,s}^2}\right)[\vv_{m}^{svd}]_{s}[\vv_{m}^{svd}]_{s}^\top,\quad {\boldsymbol{\Omega}}_{{\vv}_m^{svd}, i} := \sum_{s\in [nT]}\left(\frac{\partial \ell_{m,i,s}(\pi_{m,i,s}^*)}{\partial \pi_{m,i,s}}\right)^2[\vv_{m}^{svd}]_{s}[\vv_{m}^{svd}]_{s}^\top.
\end{equation}

The following Theorem~\ref{thm:asymp_normality_factor} establishes the asymptotic distributions for $[\widehat{\boldsymbol{\Theta}}]_{i,:}$ 
and $[\widehat{\boldsymbol{\Phi}}]_{i,:}$,
which is the basis for constructing the confidence regions for the latent positions, and also important for deriving the asymptotic distributions for the estimated connection matrices.


\begin{theorem}\label{thm:asymp_normality_factor}
Suppose Assumptions~\ref{assumption:identi}-\ref{assumption:regularities-on-ell} hold and as $n\to \infty$, $(T^{1+\epsilon}/n)\to 0$ for some small $\epsilon>0$. 
Then, for $i\in [n]$, up to diagonal sign matrices $\RR_m$, we have 
\[
	\left([\boldsymbol{\Sigma}_{\vv_1^{svd},i}^{-1}\boldsymbol{\Omega}_{\vv_1^{svd},i}\boldsymbol{\Sigma}_{\vv_1^{svd},i}^{-1}]_{S_1^*,S_1^*}\right)^{-\frac{1}{2}}\left([\widehat{\boldsymbol{\Theta}}]_{i,:}^\top-[\boldsymbol{\Theta}^*\RR_1]_{i,:}^\top\right) 
\xrightarrow{d} N(\boldsymbol{0},\II_{k_1}),
\]
\[
\left([\boldsymbol{\Sigma}_{\vv_2^{svd},i}^{-1}\boldsymbol{\Omega}_{\vv_2^{svd},i}\boldsymbol{\Sigma}_{\vv_2^{svd},i}^{-1}]_{S_2^*,S_2^*}\right)^{-\frac{1}{2}}\left([\widehat{\boldsymbol{\Phi}}]_{i,:}^\top-[\boldsymbol{\Phi}^*\RR_2]_{i,:}^\top\right)\xrightarrow{d} N(\boldsymbol{0},\II_{k_2}).
\]
Moreover, for $m\in[2]$ and $j\in [nT]$, 
\[
\left([\boldsymbol{\Sigma}_{\uu_m^{svd},j}^{-1}\boldsymbol{\Omega}_{\uu_m^{svd},j}\boldsymbol{\Sigma}_{\uu_m^{svd},j}^{-1}]_{S_m^*,S_m^*}\right)^{-\frac{1}{2}}\left([\widehat{\VV}_m^c]_{j,:}^\top - [\VV_m^{c*}\RR_m]_{j,:}^\top\right) \xrightarrow{d} N(\boldsymbol{0}, \II_{k_m}).
\]
\end{theorem}

\begin{remark}
The key challenge in establishing asymptotic normality for $[\widehat{\boldsymbol{\Theta}}]_{i,:}$ 
and $[\widehat{\boldsymbol{\Phi}}]_{i,:}$ relates to the identifiability issue discussed in Section~\ref{subsection:identifiability}.
Specifically, we can change $(\UU_m,\VV_m)$ to $(\UU_m\QQ,\VV_m(\QQ^{-1})^\top)$ for any invertible matrix $\QQ\in\R^{k_m\times k_m}$, while the value of $L_m$ remains unchanged. 
This will result in a singular Hessian matrix if we treat $L_m$ as a function of both $\UU_m$ and $\VV_m$, which makes the standard theory of M-estimation fail. To solve this, we reformulate the constraints in \eqref{op:original_OP} as a Lagrange dual problem and analyze the corresponding Hessian matrix for the Lagrangian regularized log-likelihood function. Similar techniques have been adopted in \cite{wangMaximumLikelihoodEstimation2022} and \cite{li2023statistical} under different model settings. However, the tensor structure in \eqref{eq:tensor-decomp} requires more complicated identifiability conditions as discussed in Section~\ref{subsection:identifiability} and Assumption~\ref{assumption:identi}, making it more challenging to derive the asymptotic distributions for the estimators.
\end{remark}

\begin{remark}
    Theorem~\ref{thm:asymp_normality_factor} implies that both $[\widehat\TTT]_{i,:}$ and $[\widehat\PH]_{i,:}$ are oracle estimators. Specifically, the asymptotic covariance matrix of $[\widehat\TTT]_{i,:}$, which is asymptotically equivalent to $\SG_{\vv_1^{svd},i}^{-1}$, matches with the Cramér-Rao lower bound for estimating $[\TTT^*]_{i,:}$ in the scenario when all other parameters $\{[\TTT^*]_{j,:}\}_{j\neq i}$, $\PH^*$, $\{\LLLL_t^*\}_{t=1}^T$, $\aaa^*$ and $\bbb^*$ are known. Similar property holds for $[\widehat\PH]_{i,:}$. 
\end{remark}

\begin{remark} 
The condition $T^{1+\epsilon}/n\to0$ in Theorem~\ref{thm:asymp_normality_factor} corresponds to the asymptotic regime targeted by the two node-mode unfoldings used in our main procedure, rather than an intrinsic restriction of the model. When the number of layers dominates the network size, inference can instead be based on the mode-3 unfolding. We return to this complementary regime after Theorem~\ref{thm:asymp_core_tensor}. The formal results are provided in the Supplementary Material.
\end{remark}

The asymptotic covariance matrices of $\widehat{\boldsymbol{\Theta}}$ and $\widehat{\boldsymbol{\Phi}}$ can be estimated via plug-in estimators. Define $\widehat{\uu}_m := \ve(\widehat{\UU}_m^{\top}), \widehat{\vv}_m :=\ve(\widehat{\VV}_m^{\top})$. The following Corollary~\ref{corollary:plug_in_loading} provides a guide for constructing confidence regions for $[\TTT^*]_{i,:}$ and $[\PH^*]_{i,:}$ up to sign flipping.

\begin{corollary}\label{corollary:plug_in_loading}
Under the conditions of Theorem~\ref{thm:asymp_normality_factor}, for $m \in [2]$, $i\in [n]$, and $j\in [nT]$, define $\widehat{\pi}_{m,i,j} := [\MMM_m(\widehat{\XXX})]_{i,j}$ and
\[
\widehat{\boldsymbol{\Sigma}}_{\widehat{\vv}_m,i} :=  \sum_{s \in [nT]} \left( -\frac{\partial^2 \ell_{m,i,s}(\widehat{\pi}_{m,i,s})}{\partial \pi_{m,i,s}^2} \right) [\widehat{\vv}_{m}]_{s} [\widehat{\vv}_{m}]_{s}^\top,
\quad~~
\widehat{\boldsymbol{\Omega}}_{\widehat{\vv}_m,i} := \sum_{s \in [nT]} \left( \frac{\partial \ell_{m,i,s}(\widehat{\pi}_{m,i,s})}{\partial \pi_{m,i,s}} \right)^2 [\widehat{\vv}_{m}]_{s} [\widehat{\vv}_{m}]_{s}^\top.
\]
As $n\to\infty$, $(nT)^{-1}(\widehat{\boldsymbol{\Sigma}}_{\widehat{\vv}_m, i} - \boldsymbol{\Sigma}_{\vv_m^{svd}, i}) = o_p(1)$ and $(nT)^{-1}(\widehat{\boldsymbol{\Omega}}_{\widehat{\vv}_m, i}-\boldsymbol{\Omega}_{\vv_m^{svd}, i}) = o_p(1)$. 
Let $\widehat S_m$ be the ordered index sets from Algorithm~\ref{alg:structure_recovery}. Moreover, 
\[
\left([\widehat{\boldsymbol{\Sigma}}_{\widehat{\vv}_1, i}^{-1}\widehat{\boldsymbol{\Omega}}_{\widehat{\vv}_1, i}\widehat{\boldsymbol{\Sigma}}_{\widehat{\vv}_1,i}^{-1}]_{\widehat S_1,\widehat S_1}\right)^{-\frac{1}{2}} \left([\widehat{\boldsymbol{\Theta}}]_{i,:}^\top-[\boldsymbol{\Theta}^*\RR_1]_{i,:}^\top\right) \xrightarrow{d} N(\boldsymbol{0}, \II_{k_1 \times k_1 }),
\]
\[
\left([\widehat{\boldsymbol{\Sigma}}_{\widehat{\vv}_2,i}^{-1}\widehat{\boldsymbol{\Omega}}_{\widehat{\vv}_2,i}\widehat{\boldsymbol{\Sigma}}_{\widehat{\vv}_2,i}^{-1}]_{\widehat S_2,\widehat S_2}\right)^{-\frac{1}{2}} \left([\widehat{\boldsymbol{\Phi}}]_{i,:}^\top-[\boldsymbol{\Phi}^*\RR_2]_{i,:}^\top \right) \xrightarrow{d} N(\boldsymbol{0}, \II_{k_2 \times k_2}).
\]
\end{corollary}

Corollary~\ref{corollary:plug_in_loading} enables the construction of confidence regions for the latent position estimators. It further supports inference problems such as testing whether two nodes belong to the same community and network assisted regression.


\subsection{Debiased inference for connection matrices}
\label{subsection:estimation_asymp_dist_core}

Unlike the latent position estimators, whose first-order stochastic fluctuations dominate the estimation error, the connection matrix estimator involves a fusion of two estimated loading components. As a result, second-order error propagation from the loading estimators accumulates and leads to a non-negligible asymptotic bias. This section develops a debiased inference procedure to remove this bias and establish asymptotic normality for the connection matrix estimators.
We further show how to perform hypothesis testing on whether the structures for two specific network layers are the same, which is a typical inference task for multilayer networks and has not been established for nonlinear multilayer network models yet.


\subsubsection{Debiased fusion estimator and asymptotic normality}

 Recall the connection matrix estimators in Algorithm~\ref{alg:structure_recovery}, we have
\begin{equation}
\label{eq:core_tensor_entry_estimator}
\begin{aligned}
[\widehat{\boldsymbol{\Lambda}}_{t}]_{i,j}&=\widehat{\mathcal{S}}_{i,j,t}
= \left[\mathcal{M}_1(\widehat{\mathcal{S}})\right]_{i,\, j + k_2(t - 1)} = \frac{1}{n} \sum_{s=1}^{n \cdot T} \left( [\widehat{\VV}_1^c]_{s,i} \cdot [\II_T \otimes \widehat{\mathbf{\Phi}}]_{s,\, j + k_2(t - 1)} \right)\\
&=\frac{1}{n}\sum_{s=1}^n [\widehat\VV_1^c]_{s+n(t-1), i}[\widehat\PH]_{s,j}. 
\end{aligned}
\end{equation}
To derive the asymptotic distribution for $[\widehat{\boldsymbol{\Lambda}}_{t}]_{i,j}$, we leverage the asymptotic expansion for both $[\widehat\VV_1^c]_{s+n(t-1), i}$ and $[\widehat\PH]_{s,j}$.
The key difference from standard low-rank inference is that the connection matrix estimator is not obtained directly from observations. Instead, it is reconstructed by multiplying two estimated loading matrices. Therefore, the interaction between their estimation errors generates second-order terms that are accumulated through the fusion operation.

To illustrate, we give the derivation in a high level as following, where details are deferred to the proof of Theorem~\ref{thm:asymp_core_tensor} in the appendix. Based on \eqref{eq:core_tensor_entry_estimator},
$${\footnotesize
\begin{aligned}
    &[\widehat{\boldsymbol{\Lambda}}_{t}]_{i,j} - [\RR_1{\boldsymbol{\Lambda}}^*_{t}\RR_2]_{i,j}\\ 
    =&
    \frac{1}{n}\sum_{s=1}^n \left\{ [\VV_1^{c*}\RR_1]_{s+n(t-1), i} + \Delta_{\vv_1,sit}^{(1)} + \Delta_{\vv_1,sit}^{(2)} + \text{remainder} \right\}
    \left\{ [\PH^*\RR_2]_{s, j} + \Delta_{\PH,sjt}^{(1)} + \Delta_{\PH,sjt}^{(2)} + \text{remainder} \right\} - [\RR_1{\boldsymbol{\Lambda}}^*_{t}\RR_2]_{i,j} \\
    =& \underbrace{\frac{1}{n}\sum_{s=1}^n [\PH^*\RR_2]_{s,j}\Delta_{\vv_1,sit}^{(1)}}_{\text{first order term which gives CLT}} + \underbrace{ \frac{1}{n}\sum_{s=1}^n[\PH^*\RR_2]_{s, j}\Delta_{\vv_1,sit}^{(2)} + \frac{1}{n}\sum_{s=1}^n [\VV_1^{c*}\RR_1]_{s+n(t-1),i}\Delta_{\PH,sjt}^{(2)} }_{\text{two dominating bias terms}} + o_p(\frac{1}{n}),\\
    =:& L_{ijt} + B_{1,ijt} + B_{2,ijt} + o_p(\frac{1}{n}),
\end{aligned}}
$$
where $\Delta_{\vv_1,sit}^{(1)}, \Delta_{\vv_1,sit}^{(2)}$ are the first and second order terms in the expansion of $[\widehat\VV_1^c]_{s+n(t-1), i} - [\VV_1^{c*}]_{s+n(t-1),i}$, respectively, and $\Delta_{\PH,sjt}^{(1)}, \Delta_{\PH,sjt}^{(2)}$ are the first and second order terms in the expansion of $[\widehat\PH]_{s, j} - [\PH^*]_{s, j}$, respectively.
It can be seen that only the first-order
fluctuation of $\widehat{\VV}_1^c$ contributes to the limiting variance, while the bias term comes from the aggregation of the second order terms in the error expansion of both $\widehat\VV_1^c$ and $\widehat\PH$.
In the proof of Theorem~\ref{thm:asymp_core_tensor} below, we will show that
$$
\begin{aligned}
&L_{ijt} = \sigma_{i,j,t} Z_{ijt} + o_p(\frac{1}{n}),~~\text{where}~~ Z_{ijt}\overset{d}{\to} N(0,1), \\ 
&B_{m,ijt} = b_{m,ijt} + o_p(\frac{1}{n}),~~\text{for}~~ m\in[2],
\end{aligned}
$$
where $\sigma_{i,j,t}^2$ is the asymptotic variance of $[\widehat{\boldsymbol{\Lambda}}_{t}]_{i,j}$, and $b_{1,ijt}$ and $b_{2,ijt}$ are the deterministic biases arising from the second order error terms of $\widehat\VV_1^c$ and $\widehat\PH$, respectively. 

To characterize the leading second-order bias explicitly, we impose the following additional condition on the likelihood, which is satisfied by the canonical exponential-family
likelihoods described in Remark~\ref{rk:types}.

\begin{assumption}\label{assumption:canonical_likelihood}
For each $m\in[2]$, $i\in[n]$, and $j\in[nT]$, the second derivative
$\frac{\partial^2 \ell_{m,i,j}(\pi)}{\partial\pi^2}$ is non-random for each fixed $\pi$, and $\mathbb{E}[(\frac{\partial \ell_{m,i,j}(\pi_{m,i,j}^*)}{\partial \pi})^2]=-\frac{\partial^2 \ell_{m,i,j}(\pi_{m,i,j}^*)}{\partial \pi^2}$.
\end{assumption}

\begin{remark} 
Assumption~\ref{assumption:canonical_likelihood} is used only for the second-order analysis of the connection matrix estimators and is not required for the latent-position inference in Theorem~\ref{thm:asymp_normality_factor}. It yields a tractable explicit form of the leading bias by ensuring that the second order derivative of the observed likelihood is non-random, and the Fisher identity holds.
This condition is satisfied by canonical exponential-family models, including logistic and Poisson networks. Extending the bias characterization to more general likelihoods is an interesting direction for future research.
Without this condition, additional second-order terms arising from the dependence between the first- and second-order likelihood derivatives generally need to be retained, and the corresponding bias correction requires a more involved analysis. 
\end{remark}



For $m\in[2]$, let
$\DD_m^*:=(nT)^{-1}(\VV_m^{svd})^\top\VV_m^{svd}$,
${\UU}_m^{rescale}:=\UU_m^{svd}(\DD_m^*)^{1/4}$, and
${\VV}_m^{rescale}:=\VV_m^{svd}(\DD_m^*)^{-1/4}$. This scaling gives the balanced normalization, which facilitates the estimation of the bias,
\begin{equation}\label{eq:balanced normalization}
\frac{1}{n}({\UU}_m^{rescale})^\top{\UU}_m^{rescale} = \frac{1}{nT}({\VV}_m^{rescale})^\top{\VV}_m^{rescale} = (\DD_m^*)^{\frac{1}{2}}.
\end{equation}
Further, define
${\uu}_m^{rescale}:=\operatorname{vec}(({\UU}_m^{rescale})^\top)$,
${\vv}_m^{rescale}:=\operatorname{vec}(({\VV}_m^{rescale})^\top)$,
and
${\boldsymbol{\phi}}_m^{rescale}
:=(({\uu}_m^{rescale})^\top,
({\vv}_m^{rescale})^\top)^\top$.
For the target entry $(i,j,t)$, define 
\[
{\footnotesize
\begin{aligned}
\boldsymbol{l}_{1,ijt}
:={}&
\left(
\boldsymbol{0}_{nd_1}^\top,\,
\frac{1}{n}\sum_{s=1}^n
\left[
\RR_2\PP_{S_2^*}(\DD_2^*)^{-1/4}
[{\uu}_2^{rescale}]_s
\right]_j
\left(
\ee_{n(t-1)+s}^{(nT)}
\otimes
(\DD_1^*)^{1/4}\PP_{S_1^*}^\top\ee_i^{(k_1)}
\right)^\top
\right)^\top \in\mathbb{R}^{(n+nT)d_1},
\\
\boldsymbol{l}_{2,ijt}
:={}&
\left(
\frac{1}{n}\sum_{s=1}^n
\left[
\RR_1\PP_{S_1^*}(\DD_1^*)^{1/4}
[{\vv}_1^{rescale}]_{n(t-1)+s}
\right]_i
\left(
\ee_s^{(n)}
\otimes
(\DD_2^*)^{-1/4}\PP_{S_2^*}^\top\ee_j^{(k_2)}
\right)^\top,\,
\boldsymbol{0}_{nTd_2}^\top
\right)^\top \in\mathbb{R}^{(n+nT)d_2},
\end{aligned}
}
\]
where $\PP_{S_m^*}
=
(\ee_{j_1}^{(d_m)},\ldots,\ee_{j_{k_m}}^{(d_m)})^\top$,
and $S_m^*=\{j_1,\ldots,j_{k_m}\}$ with
$j_1<\cdots<j_{k_m}$. For $m\in[2]$, let
$\boldsymbol{\omega}_{m,ijt}\in\mathbb{R}^{(n+nT)d_m}$
and
$\boldsymbol{\gamma}_{m,ijt}\in\mathbb{R}^{d_m^2}$ be the solution of the following linear equations:
\begin{equation}\label{eq:omega_gamma_def}
\begin{bmatrix}
\HH_{L_m}(\boldsymbol{\phi}_m^{rescale})
&
\CC_m
\\
\CC_m^\top
&
\boldsymbol{0}_{d_m^2\times d_m^2}
\end{bmatrix}
\begin{bmatrix}
\boldsymbol{\omega}_{m,ijt}
\\
\boldsymbol{\gamma}_{m,ijt}
\end{bmatrix}
=
\begin{bmatrix}
\boldsymbol{l}_{m,ijt}
\\
\boldsymbol{0}_{d_m^2}
\end{bmatrix}.
\end{equation}
Here, $\HH_{L_m}$ denotes the second-order derivative matrix of $L_m$, whose definition can be found in the Supplementary Material, 
and $\CC_m=\nabla_{\boldsymbol{\phi}_m}
\hh_m(\boldsymbol{\phi}_m^{rescale})
\in\mathbb{R}^{(n+nT)d_m\times d_m^2}$, where $\hh_m(\boldsymbol{\phi}_m)=\boldsymbol{0}\in\mathbb{R}^{d_m^2}$
collects the identifiability constraints of \eqref{eq:balanced normalization}.
For $i_m\in[n]$ and $i_m^\prime\in[nT]$, let
$\boldsymbol{x}_{i_m,i_m^\prime}
=
\nabla_{\boldsymbol{\phi}_m}
\pi_{m,i_m,i_m^\prime}
|_{\boldsymbol{\phi}_m=\boldsymbol{\phi}_m^{rescale}}$.
Denote $\boldsymbol{\Sigma}_{{\uu}_m^{rescale},i_m^\prime}$ analogously to~\eqref{eq:def_Sigma_Omega_u}, with $\uu_m^{svd}$ replaced by ${\uu}_m^{rescale}$ corresponding to $\boldsymbol{\phi}_m^{rescale}$. Then,
\begin{equation}\label{eq:total_leading_bias_core}
 \footnotesize
\begin{aligned}
B_{m,ijt}
&=
-\frac{1}{2}
\sum_{i_m^\prime=1}^{nT}
\tr\Bigg\{
\Bigg[
\sum_{i_m=1}^{n}
\partial_{\pi}^{3}\ell_{m,i_m,i_m^\prime}
\left(
\boldsymbol{x}_{i_m,i_m^\prime}^{\top}
\boldsymbol{\omega}_{m,ijt}
\right)
[{\uu}_m^{rescale}]_{i_m}
[{\uu}_m^{rescale}]_{i_m}^{\top}
\\[-2pt]
&\hspace{4.8cm}
+
\nabla_{[\vv_m]_{i_m^\prime}}^{2}
\left[
\boldsymbol{\gamma}_{m,ijt}^{\top}
\hh_m(\boldsymbol{\phi}_m^{rescale})
\right]
\Bigg]
\boldsymbol{\Sigma}_{{\uu}_m^{rescale},i_m^\prime}^{-1}
\Bigg\}
+
o_p\left(\frac{1}{n}\right)
\\
&=
b_{m,ijt}
+
o_p\left(\frac{1}{n}\right).
\end{aligned}
\end{equation}

\begin{remark}\label{rk:Gaussian bias}
For the Gaussian case, the bias arising from the likelihood vanishes. Furthermore, when $\boldsymbol{\Lambda}_t^*$ is diagonal, the remaining constraint-related terms cancel, such that $b_{1,ijt}+b_{2,ijt}=0$.
Further details and the corresponding derivation are provided in the Supplementary Material. This is further validated in Section~\ref{subsection:simulation_examples}.
\end{remark}

The following theorem
gives the asymptotic distribution of
$[\widehat{\boldsymbol{\Lambda}}_t]_{i,j}$.
\begin{theorem}\label{thm:asymp_core_tensor}
 Under Assumptions~\ref{assumption:identi}, \ref{assumption:regularities-on-ell} and \ref{assumption:canonical_likelihood}, for each fixed triple $(i,j,t)\in [k_1]\times [k_2]\times [T]$, as $(n,T)\to \infty$ with $(T^{1+\epsilon}/n)\to 0 $ for some small $\epsilon>0$, we have 
\[
\sigma_{i,j,t}^{-1}\left( [\widehat{\boldsymbol{\Lambda}}_t]_{i,j} - [\RR_1\boldsymbol{\Lambda}_{t}^*\RR_2]_{i,j} -(b_{1,ijt}+b_{2,ijt}) \right)
\xrightarrow{d} N \left( 0, 1 \right),
\]
where
\begin{equation*}\label{eq:core_asymp_variance}
\sigma_{i,j,t}^2 :=  \frac{1}{n^2} \sum_{s=1}^{n} 
[\PP_{S_2^*}[\uu_2^{svd}]_s]_{j}^2 \cdot 
\left( [[\boldsymbol{\Sigma}_{\uu_1^{svd},\, s+n(t - 1)}^{-1} \boldsymbol{\Omega}_{\uu_1^{svd},\, s+n(t - 1)} \boldsymbol{\Sigma}_{\uu_1^{svd},\, s+n(t - 1)}^{-1}]_{S_1^*,S_1^*}]_{i,i} \right).
\end{equation*}

\end{theorem}


Define $\widehat{\boldsymbol{\phi}}_m^{rescale} =((\widehat{\uu}_m^{rescale})^\top,
(\widehat{\vv}_m^{rescale})^\top)^\top$ as the constrained maximum likelihood estimator subject to constraints \eqref{eq:balanced normalization}.
We consider a plug-in estimator
\begin{equation}\label{eq:plugin_leading_bias_core}
\footnotesize 
\begin{aligned}
\widehat b_{m,ijt}
&=
-\frac{1}{2}
\sum_{i_m^\prime=1}^{nT}
\tr\Bigg\{
\Bigg[
\sum_{i_m=1}^{n}
\partial_{\pi}^{3}\ell_{m,i_m,i_m^\prime}
\left(
\widehat{\boldsymbol{x}}_{i_m,i_m^\prime}^{\top}
\widehat{\boldsymbol{\omega}}_{m,ijt}
\right)
[\widehat{\uu}_m^{rescale}]_{i_m}
[\widehat{\uu}_m^{rescale}]_{i_m}^{\top}
\\[-2pt]
&\hspace{4.8cm}
+
\nabla_{[\vv_m]_{i_m^\prime}}^{2}
\left[
\widehat{\boldsymbol{\gamma}}_{m,ijt}^{\top}
\hh_m(\boldsymbol{\widehat\phi}_m^{rescale})
\right]
\Bigg]
\widehat{\boldsymbol{\Sigma}}_{\widehat{\uu}_m^{rescale},i_m^\prime}^{-1}
\Bigg\}.
\end{aligned}
\end{equation}
where all functions of $\ph_m$ are evaluated at
$\widehat{\boldsymbol{\phi}}_m^{rescale}$, including $\widehat{\boldsymbol{x}}_{i_m,i_m^\prime}, \nabla_{[\vv_m]_{i_m^\prime}}^{2}
[
\widehat{\boldsymbol{\gamma}}_{m,ijt}^{\top}
\hh_m(\boldsymbol{\widehat\phi}_m^{rescale})
]$ and $\widehat{\boldsymbol{\Sigma}}_{\widehat{\uu}_m^{rescale},i_m^\prime}$.
Here, $(\widehat{\boldsymbol{\omega}}_{m,ijt},
\widehat{\boldsymbol{\gamma}}_{m,ijt})$
is obtained by solving 
\[
\begin{bmatrix}
\HH_{L_m}(\widehat{\boldsymbol{\phi}}_m^{rescale}) & \widehat{\CC}_m\\
\widehat{\CC}_m^\top & \boldsymbol{0}_{d_m^2\times d_m^2}
\end{bmatrix}
\begin{bmatrix}
\widehat{\boldsymbol{\omega}}_{m,ijt}\\
\widehat{\boldsymbol{\gamma}}_{m,ijt}
\end{bmatrix}
=
\begin{bmatrix}
\widehat{\boldsymbol{l}}_{m,ijt}\\
\boldsymbol{0}_{d_m^2}
\end{bmatrix},
\]
where $\HH_{L_m}(\widehat{\boldsymbol{\phi}}_m^{rescale})$ and $\widehat{\CC}_m$ are also evaluated at
$\widehat{\boldsymbol{\phi}}_m^{rescale}$.

\begin{remark}
Theorems~\ref{thm:asymp_normality_factor} and \ref{thm:asymp_core_tensor} focus on the regime in which the network size dominates the number of layers, with $T^{1+\epsilon}/n\to0$. A complementary inferential theory is available when the number of layers is larger. Specifically, inference can then be based on the mode-3 unfolding
\[
\mathcal{M}_3(\XXX)^\top
=(\PH\otimes\TTT)\mathcal{M}_3(\SSSS)^\top+(\II_n\otimes \mathbf{1}_n)\boldsymbol{\alpha}+(\mathbf{1}_n\otimes \II_n)\boldsymbol{\beta}.
\]
Under the corresponding mode-3 identifiability conditions, asymptotically normal inference is available for the entries of $\LLLL_t^*$ if $n^{1+\epsilon}/T\to 0$ and $T=n^{O(1)}$, and for the individual rows of $\TTT^*$ and $\PH^*$ if further $T\ll n^{3-\epsilon}$. 
Details are provided in the Supplementary Material.
\end{remark}

\begin{corollary}\label{corollary:plug_in_core}
Under the conditions of Theorem~\ref{thm:asymp_core_tensor}, the asymptotic normality result remains valid when we replace $\sigma_{i,j,t}^2$ with $\widehat{\sigma}_{i,j,t}^2$ defined by 
\begin{equation}\label{eq:core_asymp_variance2}
\widehat{\sigma}_{i,j,t}^2 = \frac{1}{ n^2}\sum_{s=1}^{n} 
[\PP_{\widehat{S}_2}[\widehat{\uu}_2]_s]_{j}^2 \cdot 
\left( [[\widehat{\boldsymbol{\Sigma}}_{\widehat{\uu}_1,\, s+n(t - 1)}^{-1} \widehat{\boldsymbol{\Omega}}_{\widehat{\uu}_1,\, s+n(t - 1)} \widehat{\boldsymbol{\Sigma}}_{\widehat{\uu}_1,\, s+n(t - 1)}^{-1}]_{\widehat{S}_1,\widehat{S}_1}]_{i,i} \right),
\end{equation}
and $b_{m,ijt}$ by $\widehat{b}_{m,ijt}$.
\end{corollary}

Corollary~\ref{corollary:plug_in_core} enables valid confidence intervals and hypothesis tests for individual entries of the layer-specific connection matrices. In the next subsection, we build on this result to test whether two network layers share the same interaction structure.

\subsubsection{Testing structural changes in multilayer networks}

In many applications, we need to test whether two network layers have the same structure. 
For example, for a dynamic network, we want to test whether there is a structural change at a specific time point. To do this, we need to partial out the effect of degree heterogeneity across different layers.
Then, based on \eqref{eq:basic_model_setup}, this is equivalent to test 
\begin{equation}\label{eq:multiple hypothesis}
H_0^{(t,t')}: \boldsymbol{\Lambda}^*_t = \boldsymbol{\Lambda}^*_{t^\prime} \quad v.s. \quad H_1^{(t,t')}: \boldsymbol{\Lambda}^*_t \neq \boldsymbol{\Lambda}^*_{t^\prime},
\end{equation}
for $t\neq t'\in[T]$. 
To test \eqref{eq:multiple hypothesis}, we first consider the following element-wise hypothesis: 
\begin{equation}\label{eq:hypothesis}
H_0^{(i,j,t,t')}:[\boldsymbol{\Lambda}^*_t]_{i,j} = [\boldsymbol{\Lambda}^*_{t^\prime}]_{i,j}\quad v.s. \quad H_1^{(i,j,t,t')}: [\boldsymbol{\Lambda}^*_t]_{i,j} \neq  [\boldsymbol{\Lambda}^*_{t^\prime}]_{i,j}
\end{equation} 
for some $(i,j)\in [k_1] \times [k_2]$ and $t\neq t^\prime \in [T]$.
To this end, we require the following corollary.
%
\begin{corollary}\label{corollary:difference_asymp}
Under the conditions of Theorem~\ref{thm:asymp_core_tensor}, for
$(i,j)\in [k_1]\times [k_2]$ and fixed $t\neq t^\prime\in [T]$, we have
\begin{equation}\label{eq:joint_core_distribution}
\begin{pmatrix}
    \widehat{\sigma}_{i,j,t}^{-1} & 0 \\
    0 & \widehat{\sigma}_{i,j,t^\prime}^{-1}
\end{pmatrix}
\begin{pmatrix}
    [\widehat{\boldsymbol{\Lambda}}_t]_{i,j}
    -
    [\RR_1\boldsymbol{\Lambda}_t^*\RR_2]_{i,j}
    -
    
    \left(
        \widehat{b}_{1,ijt}        +
        \widehat{b}_{2,ijt}
    \right)
    \\[2mm]
    [\widehat{\boldsymbol{\Lambda}}_{t^\prime}]_{i,j}
    -
    [\RR_1\boldsymbol{\Lambda}_{t^\prime}^*\RR_2]_{i,j}
    -
   
    \left(
        \widehat{b}_{1,ijt^\prime }
        +
        \widehat{b}_{2,ijt^\prime }
    \right)
\end{pmatrix}
\xrightarrow{d}
N(\boldsymbol{0},\II_2).
\end{equation}
Let
\[
\widehat{\delta}_{i,j,t,t^\prime}
:=
[\widehat{\boldsymbol{\Lambda}}_t]_{i,j}
-
[\widehat{\boldsymbol{\Lambda}}_{t^\prime}]_{i,j} -
\left\{
\left(
\widehat{b}_{1,ijt} 
+
\widehat{b}_{2,ijt} 
\right)
-
\left(
\widehat{b}_{1,ijt^\prime }
+
\widehat{b}_{2,ijt^\prime }
\right)
\right\},
\]
and $\delta_{i,j,t,t^\prime}^*:=[\RR_1\boldsymbol{\Lambda}_t^*\RR_2]_{i,j}-[\RR_1\boldsymbol{\Lambda}_{t^\prime}^*\RR_2]_{i,j}$. Then,
\begin{equation}\label{eq:difference_core_distribution}
\left(
\widehat{\sigma}_{i,j,t}^2
+
\widehat{\sigma}_{i,j,t^\prime}^2
\right)^{-\frac{1}{2}}
\left(
\widehat{\delta}_{i,j,t,t^\prime}
-
\delta_{i,j,t,t^\prime}^*
\right)
\xrightarrow{d}
N(0,1),
\end{equation}
where
$\widehat{\sigma}_{i,j,t}$ and
$\widehat{\sigma}_{i,j,t^\prime}$
are defined in \eqref{eq:core_asymp_variance2}.
\end{corollary}
By \eqref{eq:difference_core_distribution}, we could construct the confidence interval for $\delta_{i,j,t,t'}^*$ and further test \eqref{eq:hypothesis} by checking whether or not the confidence interval at specific level contains zero. 
In particular, we reject $H_0^{(i,j,t,t')}$ at level $\alpha$ if $|\widehat\delta_{i,j,t,t'}| > q_{1-\alpha/2}\left(\widehat\sigma_{i,j,t}^2 + \widehat\sigma_{i,j,t^\prime}^2\right)^{\frac{1}{2}}$, where $q_{1-\alpha/2}$ is the $1-\alpha/2$ quantile of the standard normal distribution. We could further test \eqref{eq:multiple hypothesis} by the method of Bonferroni correction. 
Specifically, we reject $H_0^{(t,t')}$ at level $\alpha$ if $|\widehat\delta_{i,j,t,t'}| > q_{1-\alpha/(2k_1k_2)}\left(\widehat\sigma_{i,j,t}^2 + \widehat\sigma_{i,j,t^\prime}^2\right)^{\frac{1}{2}}$ for some $(i,j)\in[k_1]\times [k_2]$. Alternatively, false discovery rate (FDR) control procedures can be readily applied to account for multiple testing.



\section{Numerical Experiments}\label{section:numerical_experiments}


\subsection{Simulation examples}\label{subsection:simulation_examples}

We conduct simulation studies to evaluate the finite-sample performance of our estimators and to examine the accuracy of the asymptotic normal approximations derived in Section~\ref{section:theory}.
For each simulation setting, we vary the number of nodes $n \in \{400, 800, 1200, 1600\}$ and the number of layers $T \in \{50, 100\}$. The latent dimensions are fixed at $k_1 = k_2 = 3$ and $k_\alpha = k_\beta = 2$. The parameters $\TTT^*,\PH^*,\{\LLLL_t^*\}_{t=1}^T,\UU_{\alpha}^*,\UU_{\beta}^*,\VV_{\alpha}^*,\VV_{\beta}^*$ are generated according to Assumption~\ref{assumption:identi}.

Specifically, to construct the latent positions and degree heterogeneity parameters, we first generate an $n\times (k_1 + k_\beta)$ matrix and an $n\times (k_2+k_\alpha)$ matrix with entries drawn independently from a standard Gaussian distribution. Then, $\TTT^*$ and $\PH^*$ are constructed as centered orthogonal bases by applying QR decomposition to the first $k_1$ and $k_2$ columns of the two random matrices. 
The factors $\UU_\beta^*$ and $\UU_\alpha^*$ are then constructed by projecting the remaining columns onto the orthogonal complements of $\TTT^*$ and $\PH^*$, respectively. 
Meanwhile, the matrices $\VV_\beta^*$ and $\VV_\alpha^*$ are generated independently by applying singular value decomposition to two random matrices with independent standard normal entries. 
Furthermore, for each $t\in [T]$, the connection matrix $\LLLL_t^*$ is constructed as a diagonal matrix with decreasing entries that are generated randomly from a normal distribution.



We consider three types of networks: continuous, count-valued and binary, which are generated by Gaussian, Poisson and logistic models specified in Remark~\ref{rk:types}, respectively. 
The estimation accuracies of the latent positions are evaluated by $\Delta \mathbf{\Theta} := \|\widehat{\mathbf{\TTT}}-\mathbf{\TTT}^*\RR_1\|_{2\to\infty}$ and $\Delta \mathbf{\Phi} := \|\widehat{\mathbf{\Phi}}-\mathbf{\Phi}^*\RR_2\|_{2\to\infty}$. 
To assess the asymptotic distributions for $\widehat\TTT$, $\widehat\PH$, and $\{\widehat\LLLL_t\}_{t=1}^T$, we examine three representative parameters: $[\widehat\TTT]_{1,1}$, $[\widehat\PH]_{1,1}$, and $[\widehat\LLLL_1]_{1,1}$ and report the coverage rates of the confidence intervals over 200 independent experiments.

\begin{figure}[!ht]
  \centering

  \begin{subfigure}[b]{\textwidth}
    \centering

    \begin{subfigure}[b]{0.32\textwidth}\centering 
      \includegraphics[
        width=\textwidth
      ]{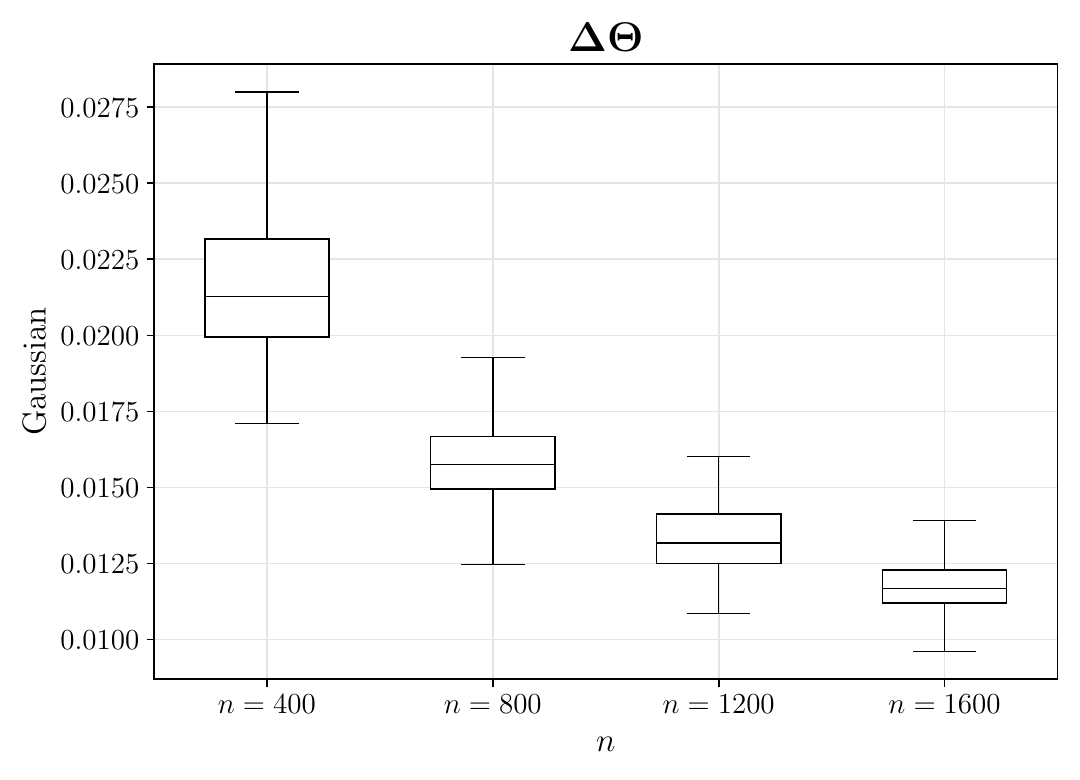}
    \end{subfigure}%
    \begin{subfigure}[b]{0.32\textwidth}\centering 
      \includegraphics[
        width=\textwidth
      ]{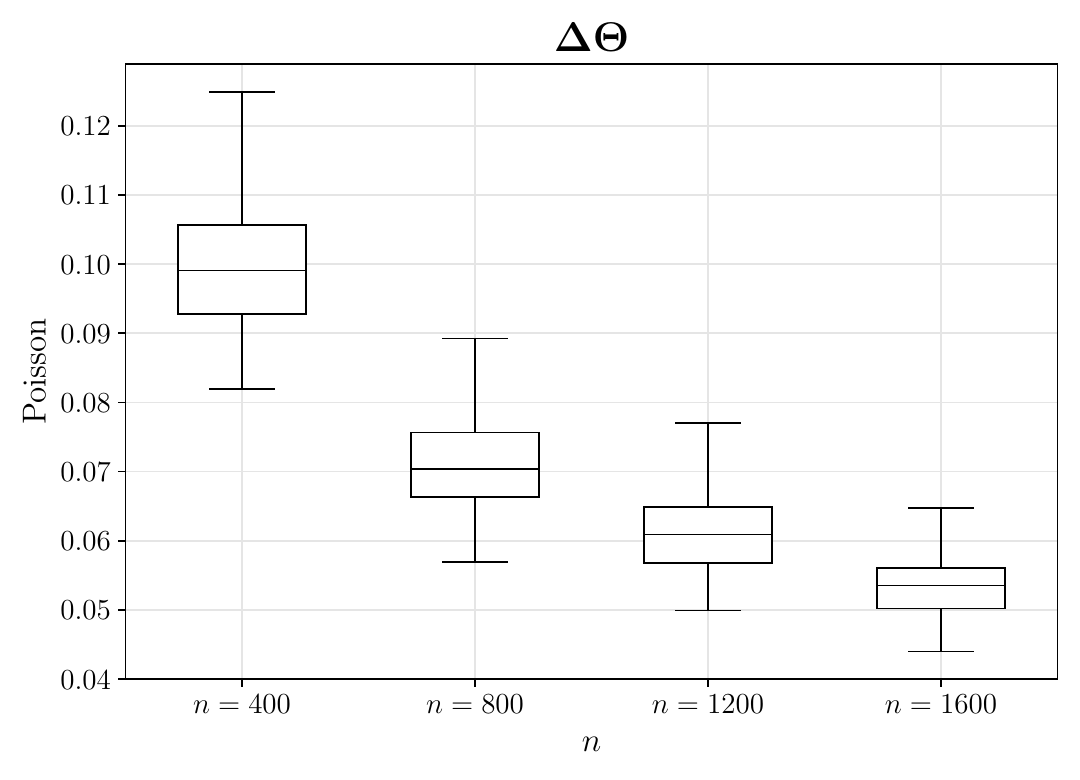}
    \end{subfigure}%
    \begin{subfigure}[b]{0.32\textwidth}\centering
      \includegraphics[
        width=\textwidth
      ]{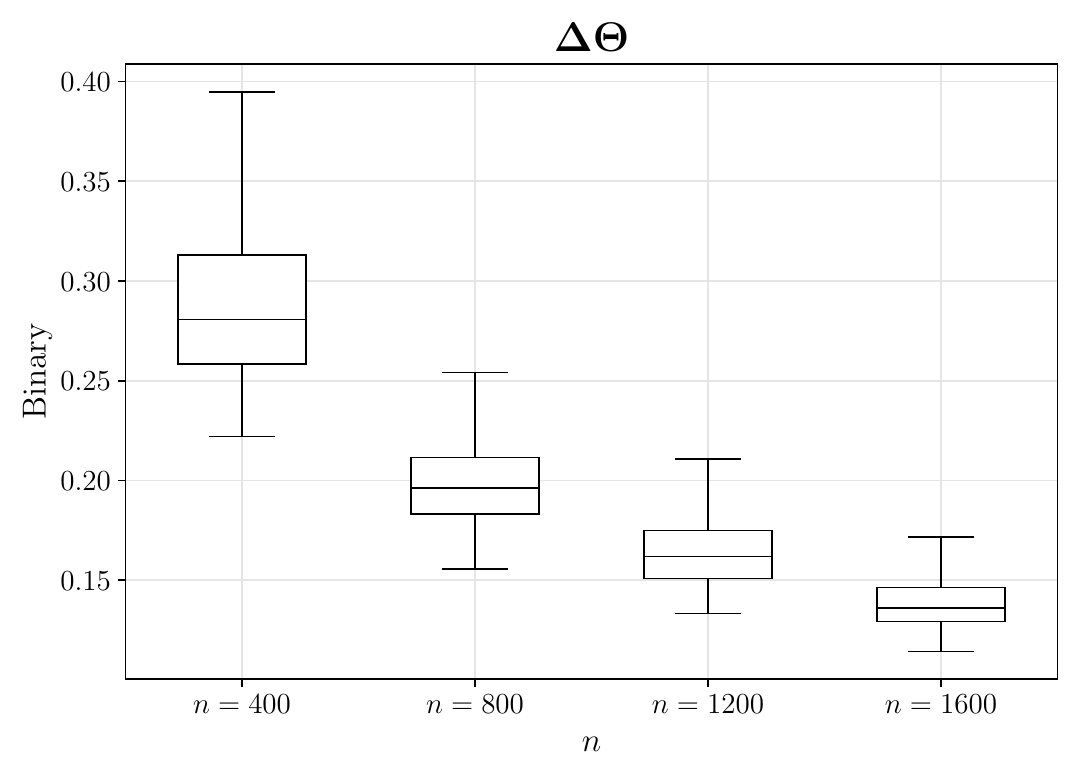}
    \end{subfigure}

    \vspace{-0.3em}

    \begin{subfigure}[b]{0.32\textwidth}\centering 
      \includegraphics[
        width=\textwidth
      ]{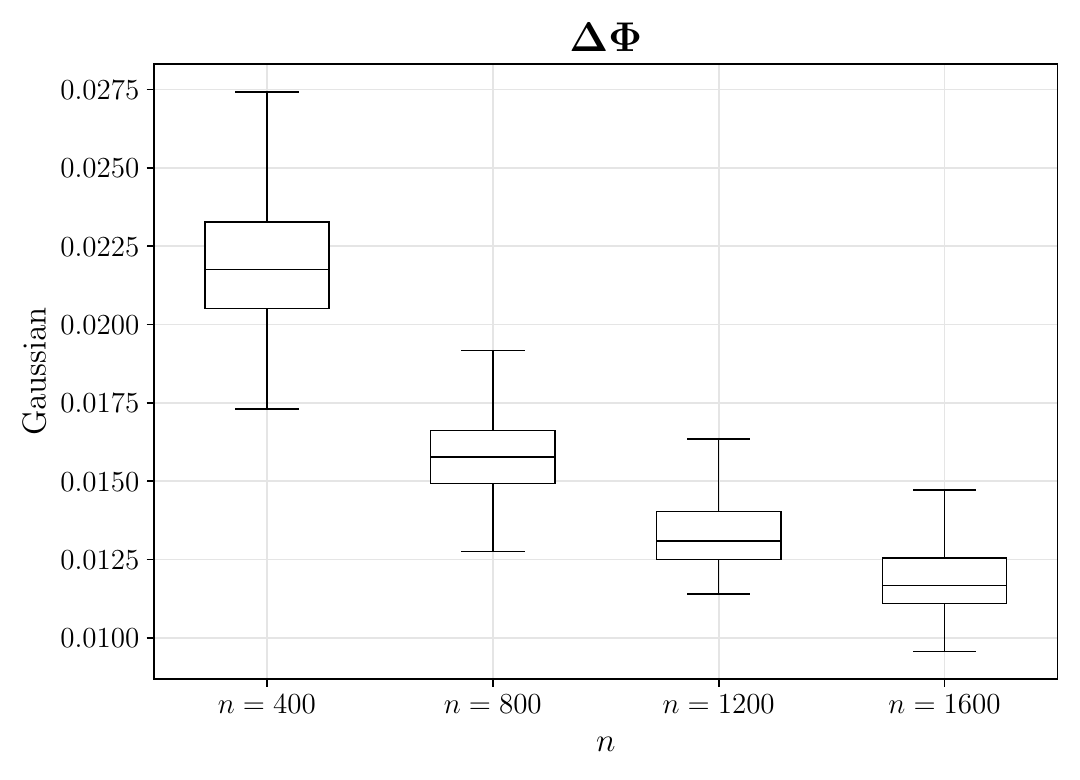}
    \end{subfigure}%
    \begin{subfigure}[b]{0.32\textwidth}\centering 
      \includegraphics[
        width=\textwidth
      ]{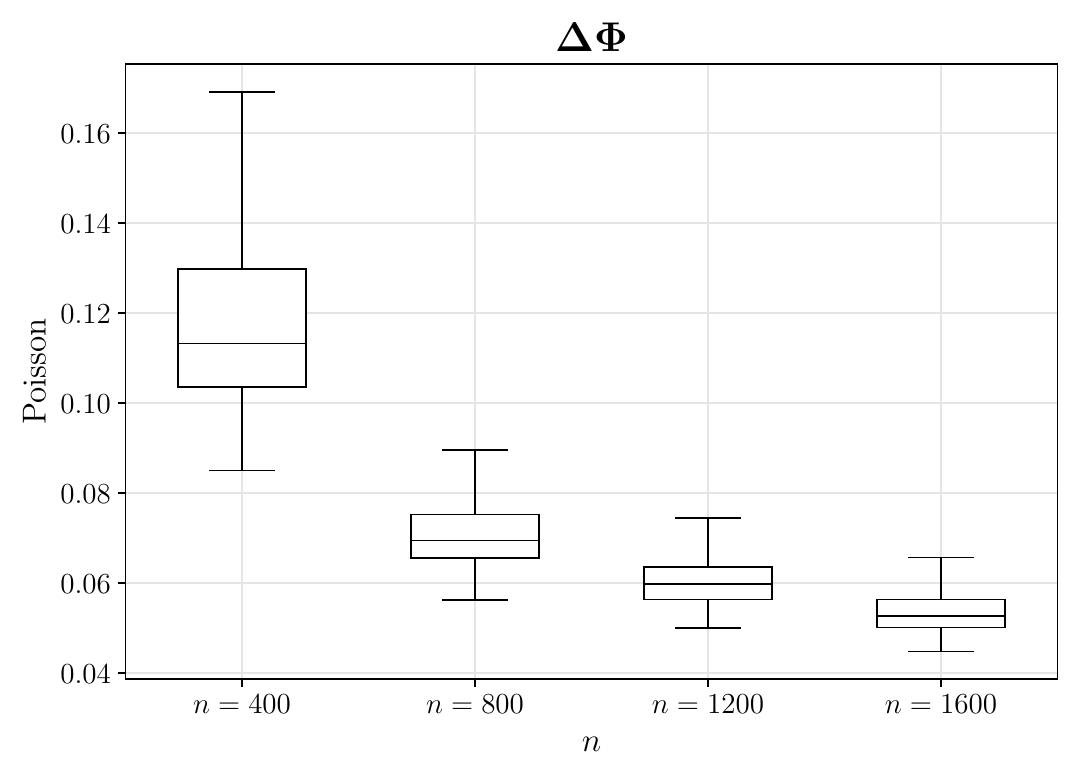}
    \end{subfigure}%
    \begin{subfigure}[b]{0.32\textwidth}\centering
      \includegraphics[
        width=\textwidth
      ]{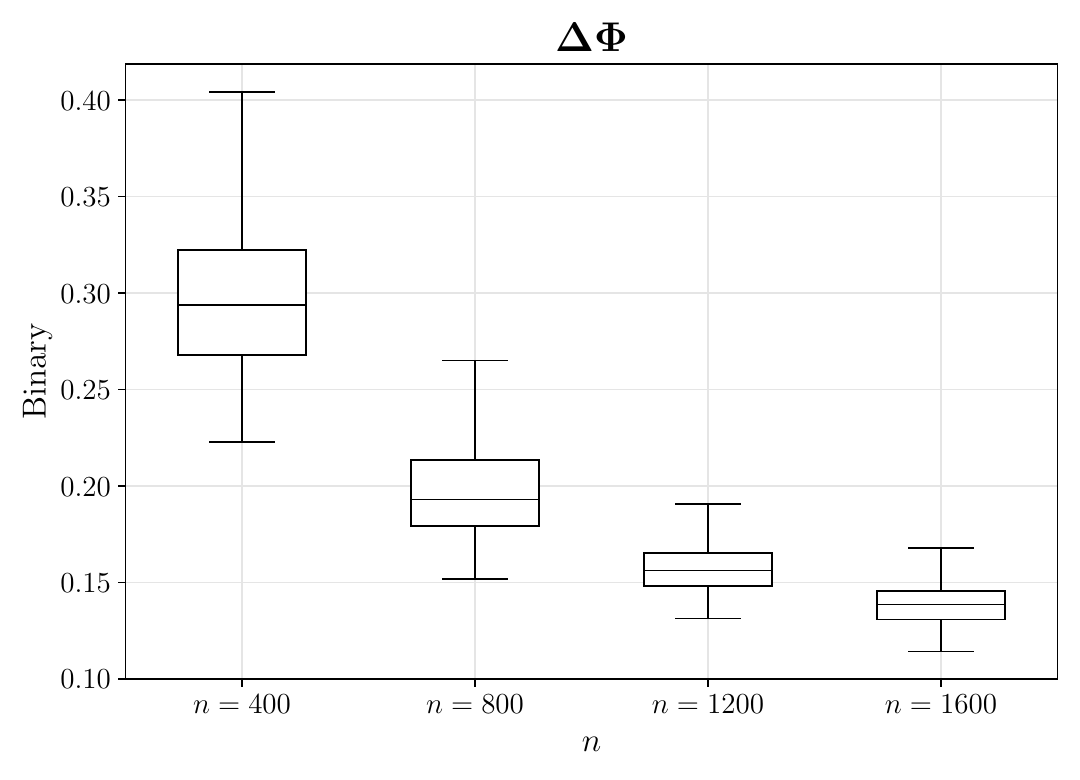}
    \end{subfigure}

    \caption{$T = 50$}
    \label{fig:boxplot-t50}
    \vspace{-0.3em}
  \end{subfigure}

  \begin{subfigure}[b]{\textwidth}
    \centering

    \begin{subfigure}[b]{0.32\textwidth}\centering 
      \includegraphics[
        width=\textwidth
      ]{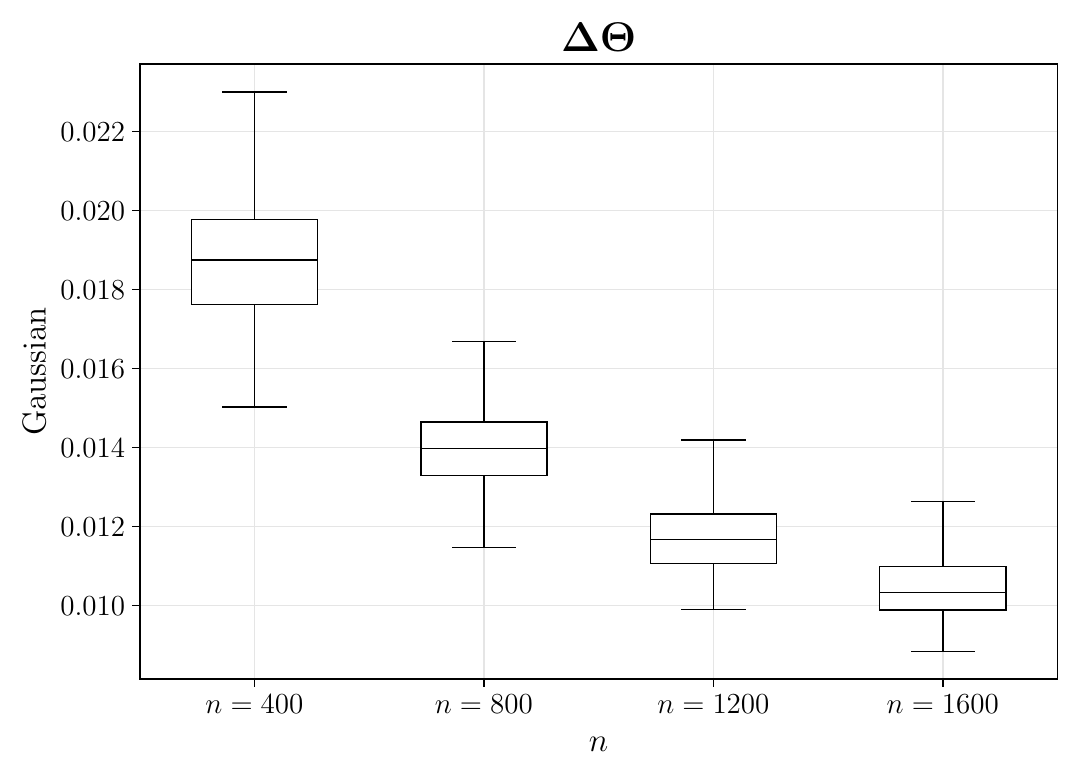}
    \end{subfigure}%
    \begin{subfigure}[b]{0.32\textwidth}\centering 
      \includegraphics[
        width=\textwidth
      ]{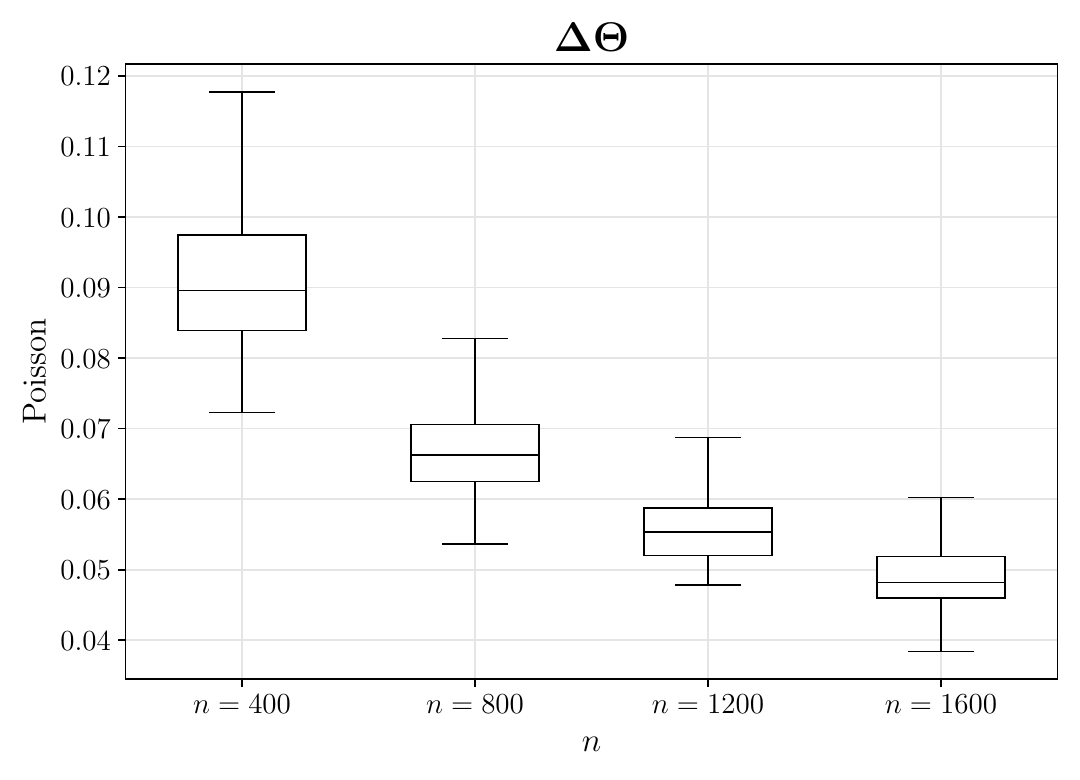}
    \end{subfigure}%
    \begin{subfigure}[b]{0.32\textwidth}\centering
      \includegraphics[
        width=\textwidth
      ]{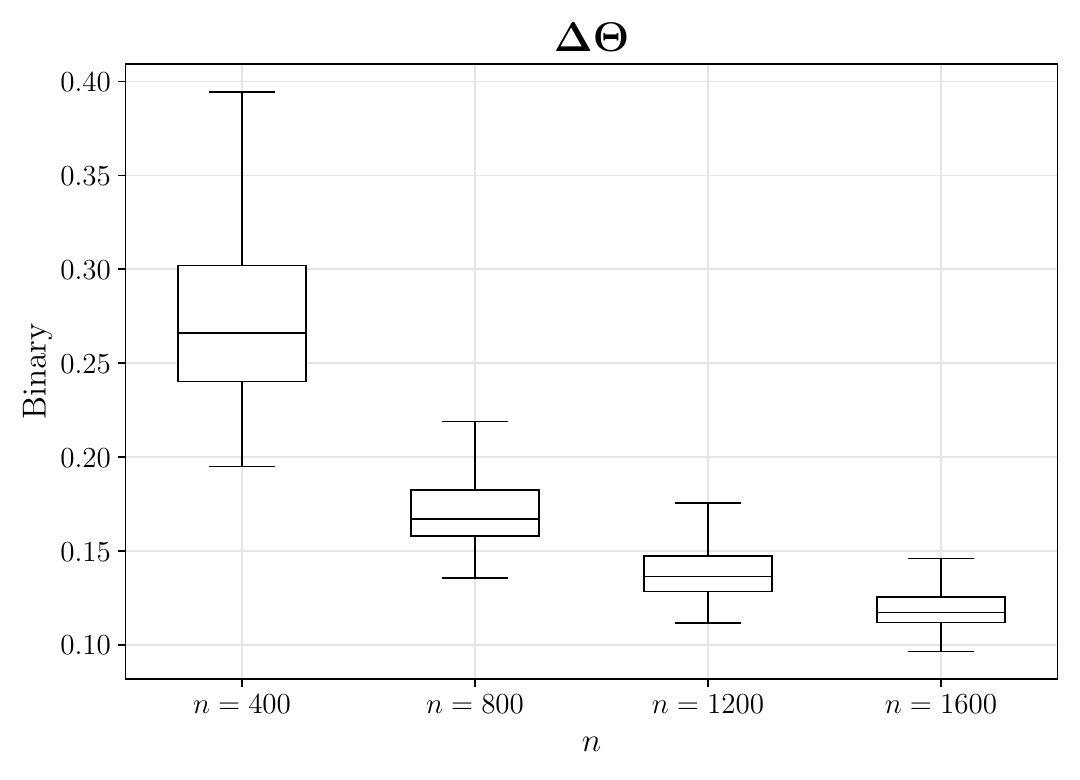}
    \end{subfigure}

    \vspace{-0.3em}

    \begin{subfigure}[b]{0.32\textwidth}\centering 
      \includegraphics[
        width=\textwidth
      ]{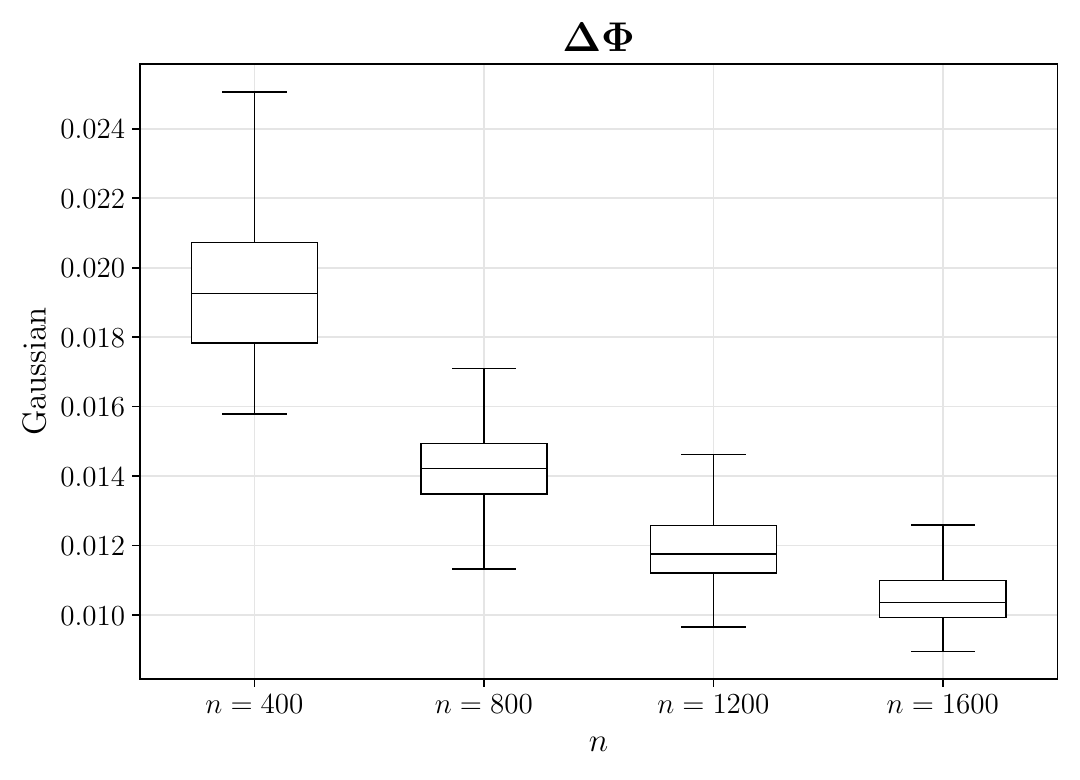}
    \end{subfigure}%
    \begin{subfigure}[b]{0.32\textwidth}\centering 
      \includegraphics[
        width=\textwidth
      ]{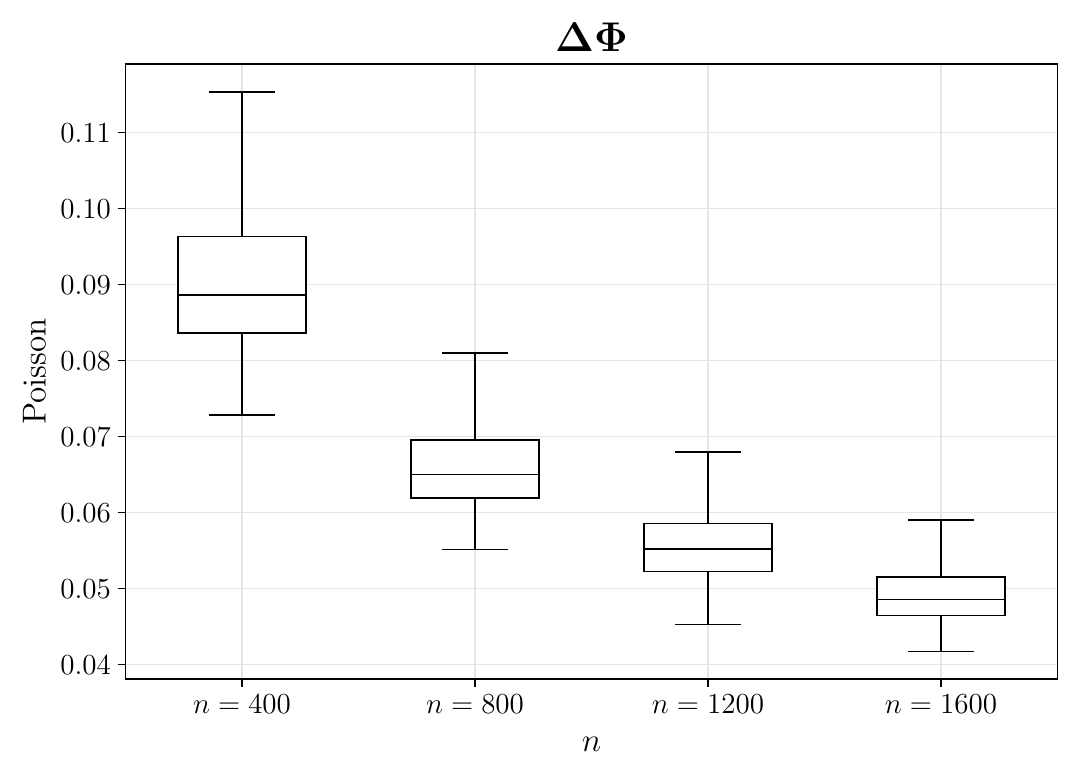}
    \end{subfigure}%
    \begin{subfigure}[b]{0.32\textwidth}\centering
      \includegraphics[
        width=\textwidth
      ]{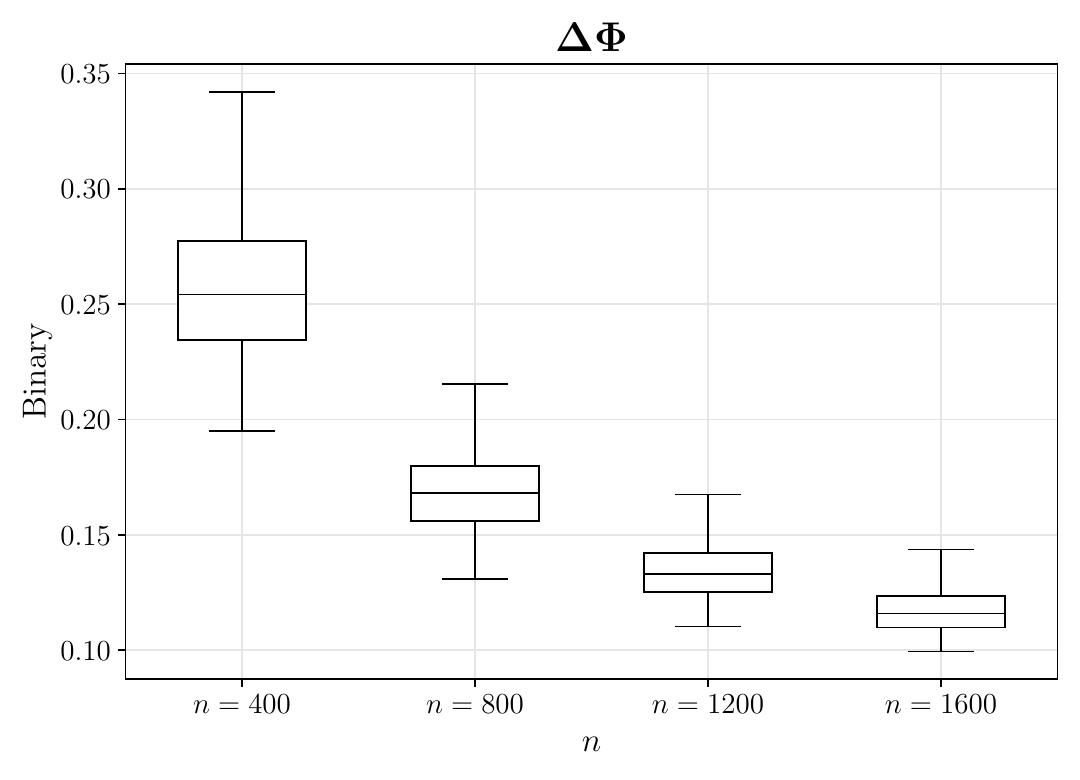}
    \end{subfigure}

    \vspace{-0.3em}
    \caption{$T = 100$}
    \label{fig:boxplot-t100}
  \end{subfigure}

  \caption{
  Boxplots for the estimation errors, $\Delta\TTT$ and $\Delta\PH$, over 200 independent experiments under different settings.
  Columns correspond to Gaussian, Poisson, and Binary settings, while rows correspond to $\boldsymbol{\Theta}$ and $\boldsymbol{\Phi}$.
  }
  \label{fig:boxplot-grid}
\end{figure}

\begin{figure}[!ht]
  \centering

  \begin{subfigure}{\textwidth}
    \centering
    \includegraphics[width=0.32\textwidth]{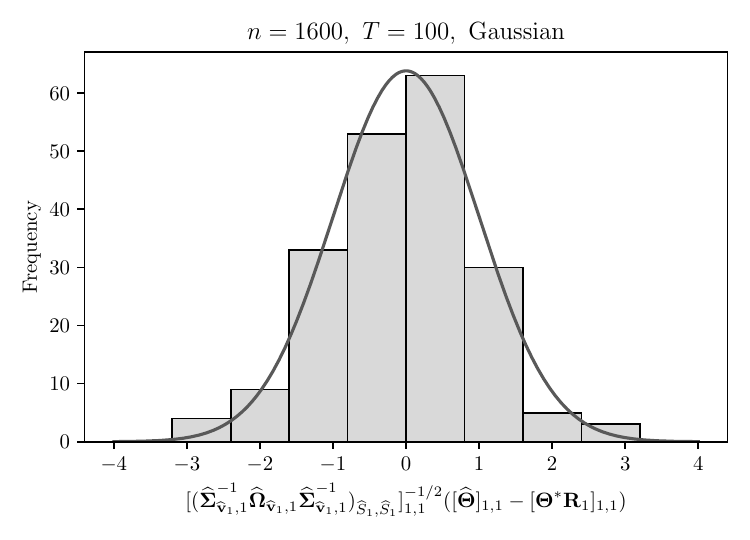}
    \includegraphics[width=0.32\textwidth]{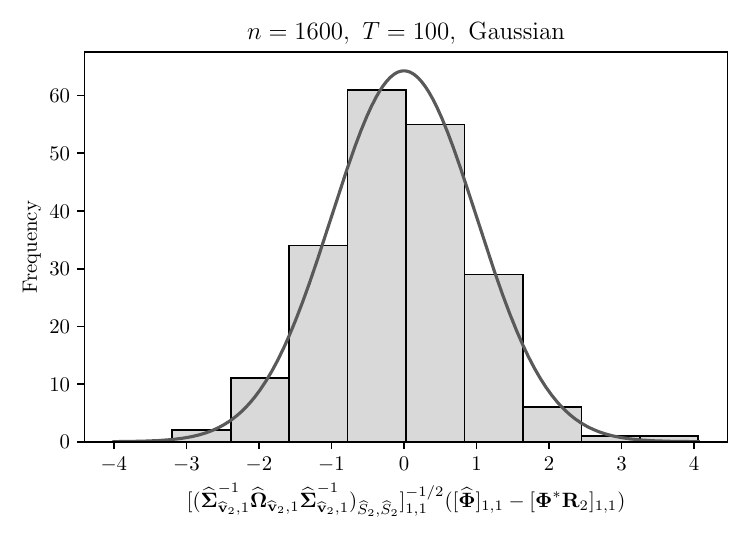}
    \includegraphics[width=0.32\textwidth]{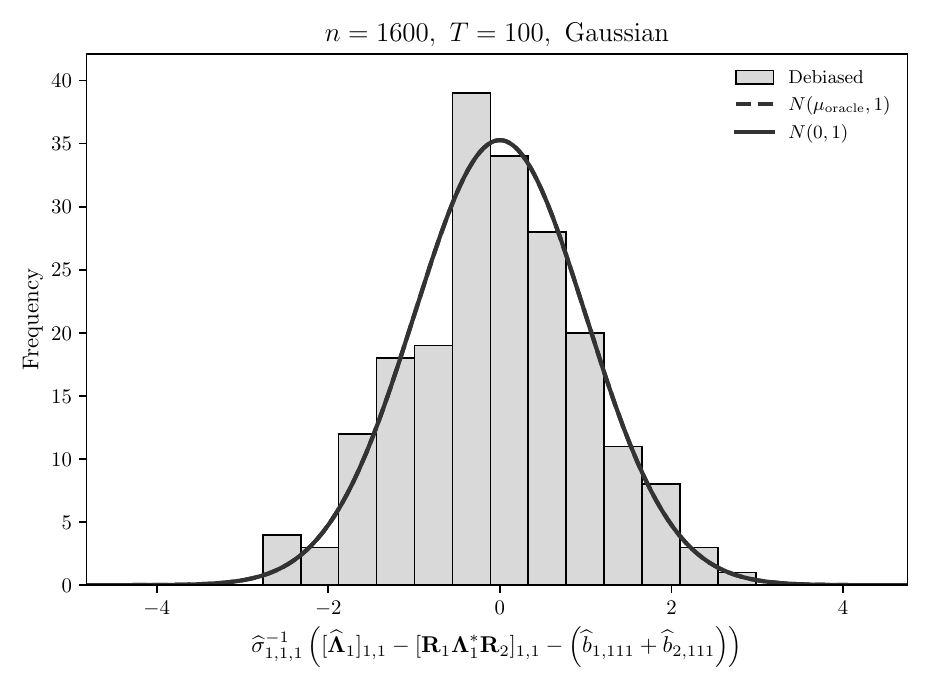}
    \caption{Gaussian}
  \end{subfigure}

  \vspace{0.6em}

  \begin{subfigure}{\textwidth}
    \centering
    \includegraphics[width=0.32\textwidth]{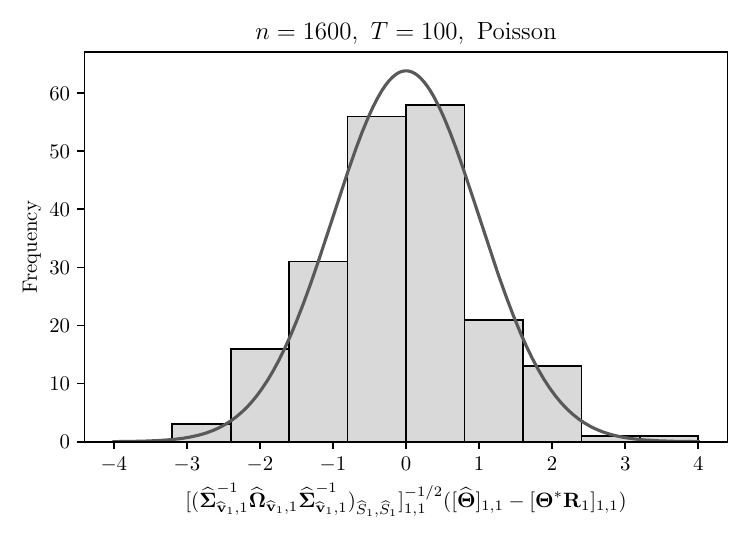}
    \includegraphics[width=0.32\textwidth]{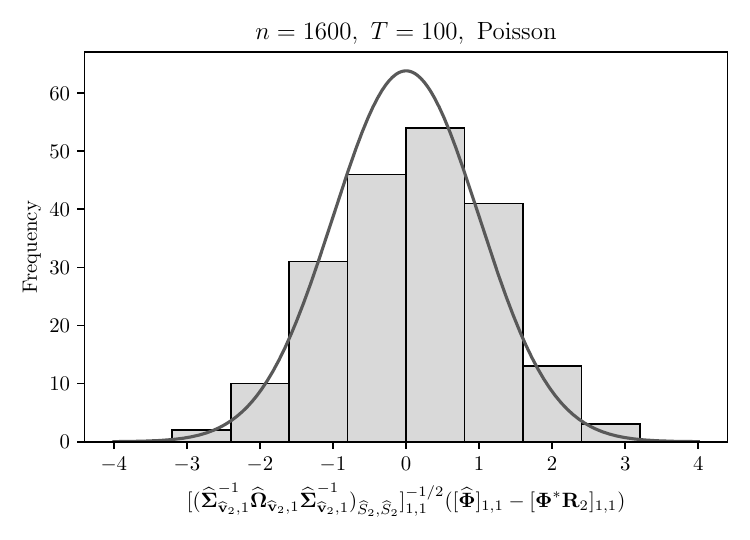}
    \includegraphics[width=0.32\textwidth]{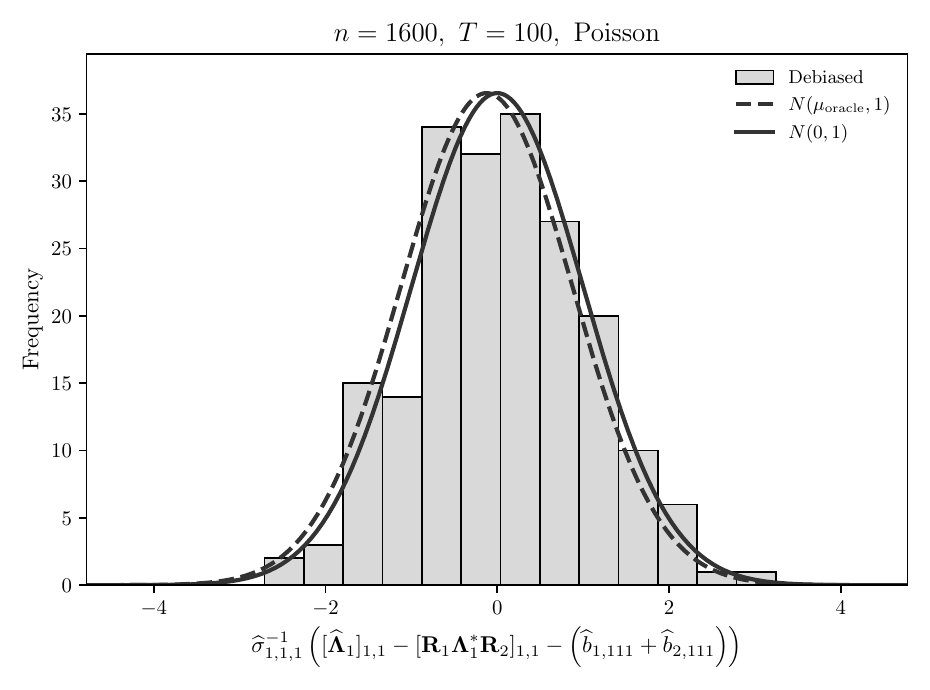}
    \caption{Poisson}
  \end{subfigure}

  \vspace{0.6em}

  \begin{subfigure}{\textwidth}
    \centering
    \includegraphics[width=0.32\textwidth]{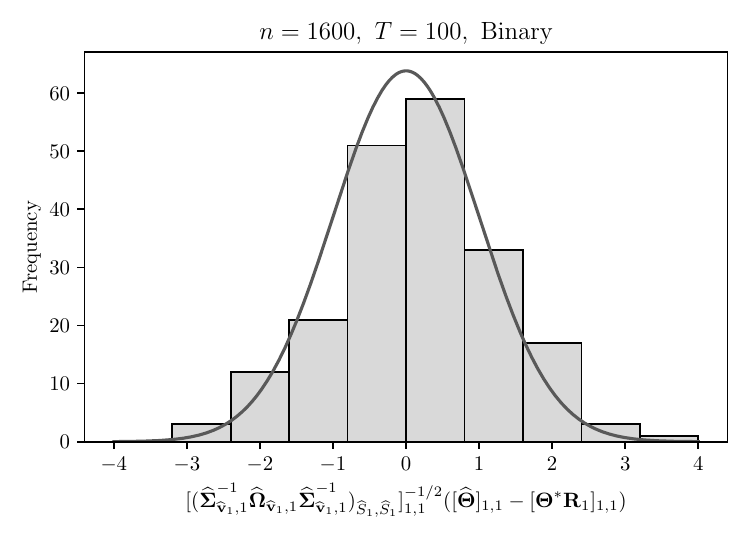}
    \includegraphics[width=0.32\textwidth]{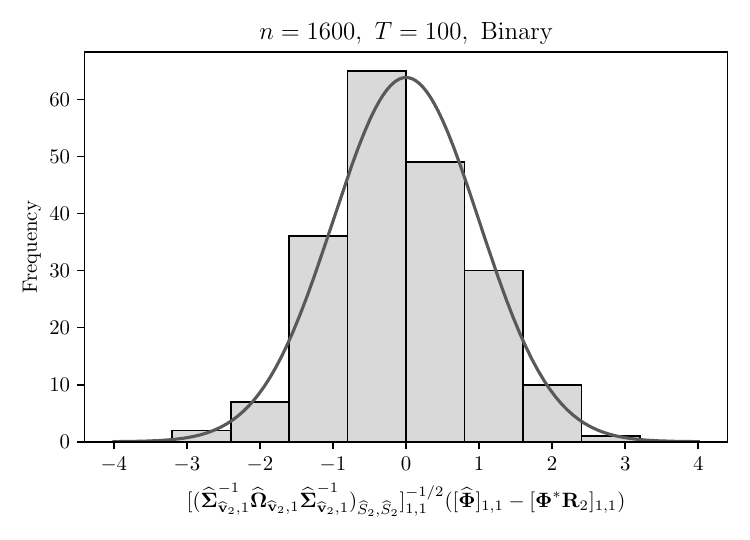}
    \includegraphics[width=0.32\textwidth]{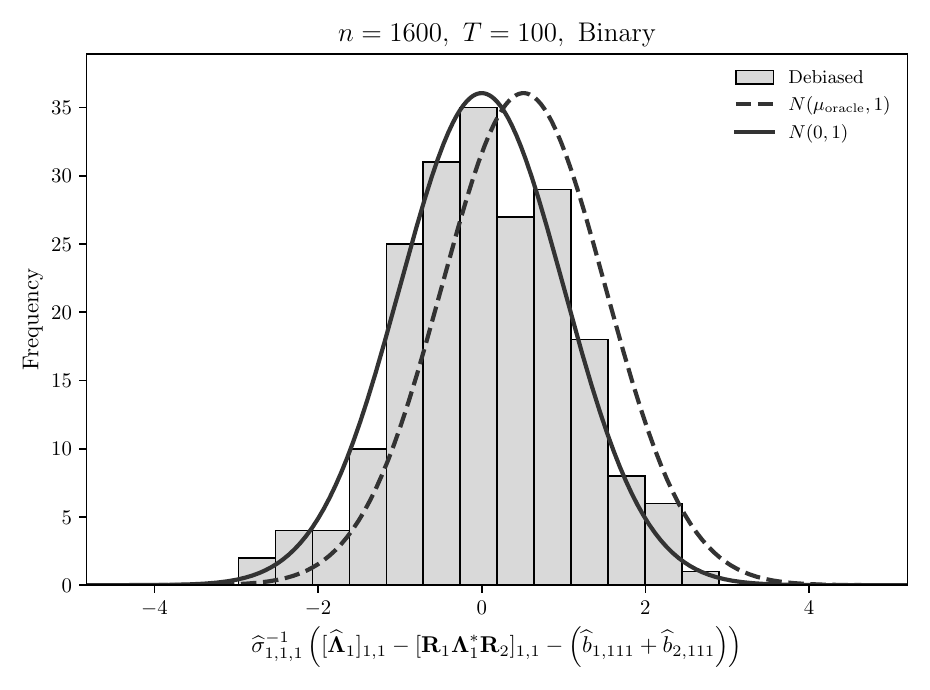}
    \caption{Binary}
  \end{subfigure}
\caption{
Histograms for studentized versions of the loading errors
$[\widehat\TTT]_{1,1}-[\TTT^*\RR_1]_{1,1}$,
$[\widehat\PH]_{1,1}-[\PH^*\RR_2]_{1,1}$, and
studentized versions of the debiased connection matrix errors
$[\widehat\LLLL_1]_{1,1}-[\RR_1\LLLL_1^*\RR_2]_{1,1} - (\widehat b_{1,111} + \widehat b_{2,111})$ under $n=1600$ and $T=100$. The first two columns compare the empirical distributions with the standard normal density. 
The third column displays the empirical distributions with the standard normal densities (solid) and the biased normal densities (dashed).
}
  \label{fig:hist_3x3_100_1600}
\end{figure}

The results are summarized in Figure~\ref{fig:boxplot-grid},
Figure~\ref{fig:hist_3x3_100_1600}, and
Table~\ref{tab:coverage-merged}. 
Figure~\ref{fig:boxplot-grid}
shows the estimation errors of $\widehat\TTT$ and $\widehat\PH$ under different settings. The errors
decrease as $n$ and $T$ increase, and the empirical behavior is
consistent with the convergence rates in
Theorem~\ref{thm:loading_individual_consistency}. 


Table~\ref{tab:coverage-merged} reports the coverage rates of nominal
$95\%$ confidence intervals over 200 independent experiments. For the
loading entries $[\TTT^*]_{1,1}$ and $[\PH^*]_{1,1}$, the empirical
coverages are generally close to the nominal level for moderate and
large $n$. This supports the plug-in normal approximation in
Corollary~\ref{corollary:plug_in_loading}. For the connection matrix
entry $[\LLLL_1^*]_{1,1}$, the comparison between the raw and debiased
intervals illustrates the necessity of the bias correction in
Theorem~\ref{thm:asymp_core_tensor} and
Corollary~\ref{corollary:plug_in_core}. Under the Gaussian model, the
raw and debiased procedures perform almost identically, which is consistent with Remark~\ref{rk:Gaussian bias}.
Under the Poisson model, the debiased intervals
generally move the empirical coverage slightly closer to the nominal level,
especially when $T=100$. The improvement is most pronounced under the
binary model: the raw fusion intervals systematically undercover,
whereas the debiased intervals substantially improve the coverage and
are much closer to the target $95\%$ level.

Figure~\ref{fig:hist_3x3_100_1600} further visualizes these findings
for $n=1600$ and $T=100$. The first two columns show that the studentized empirical
distributions of
$[\widehat\TTT-\TTT^*\RR_1]_{1,1}$ and
$[\widehat\PH-\PH^*\RR_2]_{1,1}$ are well approximated by the standard normal distribution. 

The third column displays the studentized debiased connection-matrix errors. For comparison, we overlay both the standard normal density, corresponding to the debiased limit, and the normal density shifted by the theoretical standardized bias $(b_{1,111} + b_{2,111})/\sigma_{1,1,1}$.
In all three settings, the histograms for the studentized debiased errors are well approximated by $N(0,1)$. The separation between the standard normal and the bias-shifted normal approximation is visible for the Poisson and, more prominently, the binary models, whereas the two curves essentially coincide in the Gaussian case.
These results confirm that the
second-order correction is not merely a theoretical refinement, but is
important for valid finite-sample inference on the layer-specific
connection matrices.

\begin{table}[!ht]
  \centering
  \scriptsize
  \begin{subtable}[t]{\textwidth}
    \centering
    \begin{tabular}{llcccc}
      \toprule
      \textbf{Parameter} & \textbf{Setting} & $n=400$ & $n=800$ & $n=1200$ & $n=1600$ \\
      \midrule
      \multirow{3}{*}{$[\mathbf{\Theta}]_{1,1}$} & \textbf{Gaussian} & 0.950 (0.0154) & 0.955 (0.0147) & 0.915 (0.0197) & 0.960 (0.0139) \\
       & \textbf{Poisson} & 0.860 (0.0245) & 0.950 (0.0154) & 0.920 (0.0192) & 0.955 (0.0147) \\
       & \textbf{Binary} & 0.905 (0.0207) & 0.910 (0.0202) & 0.915 (0.0197) & 0.920 (0.0192) \\
      \midrule
      \multirow{3}{*}{$[\mathbf{\Phi}]_{1,1}$} & \textbf{Gaussian} & 0.975 (0.0110) & 0.955 (0.0147) & 0.960 (0.0139) & 0.950 (0.0154) \\
       & \textbf{Poisson} & 0.935 (0.0174) & 0.930 (0.0180) & 0.950 (0.0154) & 0.960 (0.0139) \\
       & \textbf{Binary} & 0.980 (0.0099) & 0.940 (0.0168) & 0.925 (0.0186) & 0.950 (0.0154) \\
      \midrule
      \multirow{3}{*}{$[\LLLL_1]_{1,1}$} & \textbf{Gaussian} & 0.955 (0.0147) & 0.925 (0.0186) & 0.950 (0.0154) & 0.950 (0.0154) \\
       & \textbf{Poisson} & 0.930 (0.0180) & 0.960 (0.0139) & 0.955 (0.0147) & 0.955 (0.0147) \\
       & \textbf{Binary} & 0.920 (0.0192) & 0.900 (0.0212) & 0.890 (0.0221) & 0.915 (0.0197) \\
      \midrule
      \multirow{3}{*}{$[\LLLL_1^{\text{debias}}]_{1,1}$} & \textbf{Gaussian} & 0.955 (0.0147) & 0.925 (0.0186) & 0.950 (0.0154) & 0.950 (0.0154) \\
       & \textbf{Poisson} & 0.940 (0.0168) & 0.960 (0.0139) & 0.955 (0.0147) & 0.965 (0.0130) \\
       & \textbf{Binary} & 0.975 (0.0110) & 0.970 (0.0121) & 0.955 (0.0147) & 0.960 (0.0139) \\
      \bottomrule
    \end{tabular}
    \caption{$T = 50$}
    \label{tab:coverage-plugin-T50}
  \end{subtable}

  \bigskip

  \begin{subtable}[t]{\textwidth}
    \centering
    \begin{tabular}{llcccc}
      \toprule
      \textbf{Parameter} & \textbf{Setting} & $n=400$ & $n=800$ & $n=1200$ & $n=1600$ \\
      \midrule
      \multirow{3}{*}{$[\mathbf{\Theta}]_{1,1}$} & \textbf{Gaussian} & 0.970 (0.0121) & 0.970 (0.0121) & 0.945 (0.0161) & 0.940 (0.0168) \\
       & \textbf{Poisson} & 0.930 (0.0180) & 0.945 (0.0161) & 0.945 (0.0161) & 0.925 (0.0186) \\
       & \textbf{Binary} & 0.865 (0.0242) & 0.925 (0.0186) & 0.925 (0.0186) & 0.910 (0.0202) \\
      \midrule
      \multirow{3}{*}{$[\mathbf{\Phi}]_{1,1}$} & \textbf{Gaussian} & 0.935 (0.0174) & 0.950 (0.0154) & 0.980 (0.0099) & 0.955 (0.0147) \\
       & \textbf{Poisson} & 0.940 (0.0168) & 0.905 (0.0207) & 0.945 (0.0161) & 0.945 (0.0161) \\
       & \textbf{Binary} & 0.835 (0.0262) & 0.915 (0.0197) & 0.930 (0.0180) & 0.945 (0.0161) \\
      \midrule
      \multirow{3}{*}{$[\LLLL_1]_{1,1}$} & \textbf{Gaussian} & 0.945 (0.0161) & 0.955 (0.0147) & 0.950 (0.0154) & 0.950 (0.0154) \\
       & \textbf{Poisson} & 0.915 (0.0197) & 0.950 (0.0154) & 0.945 (0.0161) & 0.950 (0.0154) \\
       & \textbf{Binary} & 0.930 (0.0180) & 0.905 (0.0207) & 0.915 (0.0197) & 0.905 (0.0207) \\
      \midrule
      \multirow{3}{*}{$[\LLLL_1^{\text{debias}}]_{1,1}$} & \textbf{Gaussian} & 0.945 (0.0161) & 0.955 (0.0147) & 0.950 (0.0154) & 0.950 (0.0154) \\
       & \textbf{Poisson} & 0.940 (0.0168) & 0.970 (0.0121) & 0.950 (0.0154) & 0.960 (0.0139) \\
       & \textbf{Binary} & 0.955 (0.0147) & 0.950 (0.0154) & 0.945 (0.0161) & 0.925 (0.0186) \\
      \bottomrule
    \end{tabular}
    \caption{$T = 100$}
    \label{tab:coverage-plugin-T100}
  \end{subtable}
  \caption{Coverage rates of 95\% confidence intervals for $[\mathbf{\Theta}^*]_{1,1}$ and $[\mathbf{\Phi}^*]_{1,1}$ and $[\LLLL_1^*]_{1,1}$ over 200 independent experiments under different settings and $n,T$. Standard errors are reported in parentheses.}
\label{tab:coverage-merged}
\end{table}

\subsection{Real data}\label{subsection:real_data}

The Correlates Of War \citep[COW,][]{barbieri_keshk_cow_trade_2016} is a comprehensive project containing a variety of datasets capturing international behavior over the years. 
We apply the proposed method to analyze the trade dataset from COW, which provides bilateral import value for 207 countries and districts from 1870 to 2014. A subset of 100 active countries is selected from 1965 to 2014 based on average annual trade volume.
To reduce the strong right skewness of trade volumes while retaining zero-valued observations, we transform the import values as $\widetilde y_{ijt}=\log(y_{ijt}+10^{-8})$. 
This leads to a multilayer network with $n=100$ nodes and $T = 50$ layers, as well as an adjacency tensor $\YYY = (\widetilde y_{ijt})_{100\times 100\times 50}$ with continuous entries. We apply the Gaussian model such that $\YYY = \XXX + \EEEE$ with each $\EEEE_{i,j,t}$ following a Gaussian distribution with mean 0 and unknown variance $\sigma_0^2$.

In order to determine the dimensions of the latent positions, we examine the singular values of the two-sided centered data matrices $\JJ_{n}\MMM_m(\YYY) \JJ_{n, T}^\top,m\in [2]$. The scree plots in the Supplementary Material reveal a clear elbow at both the third singular values, and thus we choose $k_1 = k_2 = 3$. 
We set $k_\alpha=k_\beta=1$ to obtain a parsimonious low-rank representation of the layer-varying degree effects.

\begin{figure}[!htbp]
\centering
\begin{subfigure}[t]{0.48\textwidth} 
  \centering
  \includegraphics[width=\linewidth, height=6.7cm]{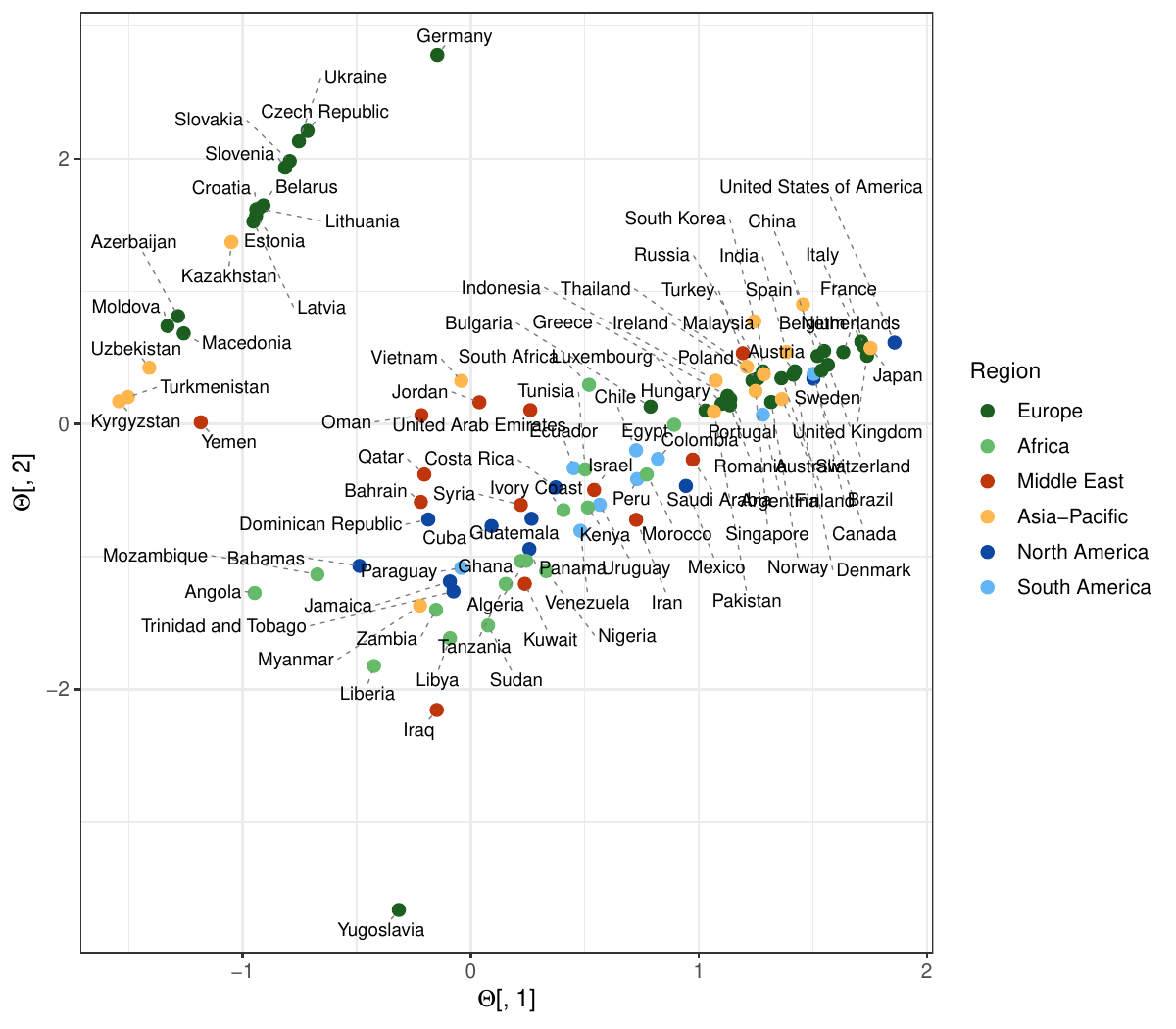} 
  \caption{$[\widehat\TTT]_{:,1}$ and $[\widehat\TTT]_{:,2}$}
  \label{fig:theta12full}
\end{subfigure}
\hfill 
\begin{subfigure}[t]{0.48\textwidth}
  \centering
  \includegraphics[width=\linewidth, height=6.7cm]{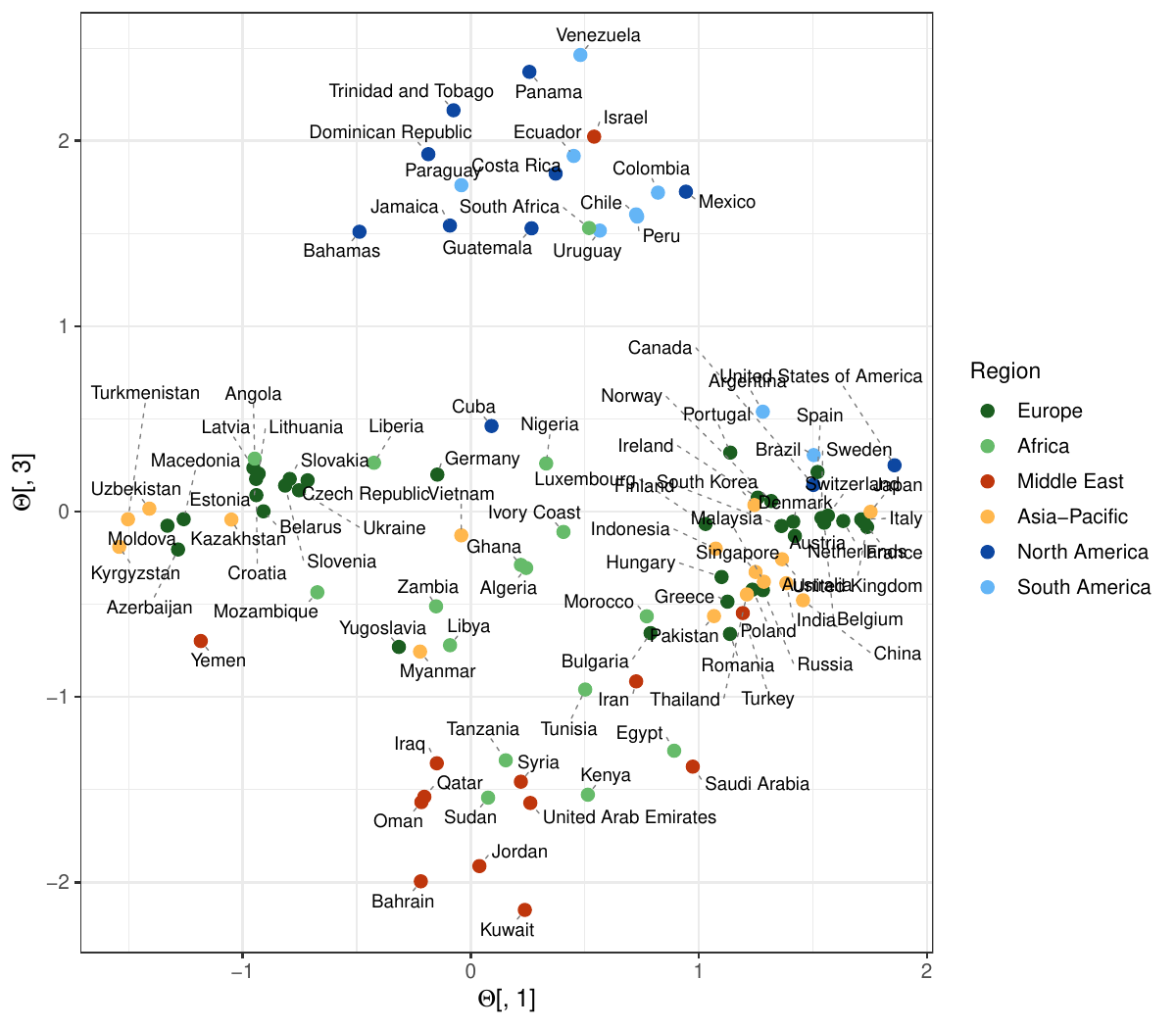}
  \caption{$[\widehat\TTT]_{:,1}$ and $[\widehat\TTT]_{:,3}$}
  \label{fig:theta13}
\end{subfigure}

\caption{Visualizations of the $\{\widehat\ttt_i\}_{i=1}^n$. Panel (a) shows the first two dimensions and panel (b) shows the first and third dimensions. Countries are colored according to region.}
\label{fig:theta_panels}
\end{figure}

Countries with large entries in $[\widehat\TTT]_{:,1}$, as shown in Figure~\ref{fig:theta_panels}, are mostly large, high-income, industrialized economies. Meanwhile, countries with low values are mostly peripheral exporters that have weaker integration into the global trade network. Hence, $[\widehat{\TTT}]_{:,1}$ reflects the degree of ``global-hubness'' of the exporters, it distinguishes core industrialized economies (global hubs) from peripheral, less integrated exporters. As for the second component of $\widehat{\mathbf{\Theta}}$, it seems to assign high values on emerging economies.
As shown in Figure~\ref{fig:theta12full}, top countries are largely post-Soviet countries, with China and South Korea also appearing as rising economic power over the past decades. At the other end of the spectrum, the country with the lowest score is Yugoslavia, a country that dissolved in the early 1990s. Overall, this component highlights patterns specific to emerging economies. 
Figure~\ref{fig:theta13} reveals a strong pattern of geographical clustering: countries from the same region tend to take on similar values of $[\widehat{\TTT}]_{:,3}$, countries in North and South America cluster toward the upper part of the plot, whereas Middle-Eastern countries gather near the bottom.
The three components of $\widehat{\PH}$ exhibit patterns similar to those of $\widehat{\TTT}$. The first component loads heavily on  large, developed economies with strong consumption demand, while the second component assigns high values to emerging markets. We can also observe a degree of geographical clustering on $[\widehat{\PH}]_{:,3}$. 
The visualization of $\{[\widehat{\PH}]_{i:,}\}_{i=1}^n$, as well as the top and bottom 15 countries in each of the three dimensions, are provided in the Supplementary Material.

\begin{figure}[!t]
\centering

\begin{subfigure}[t]{0.48\textwidth}
  \centering
  \includegraphics[width=\linewidth, height=4cm]{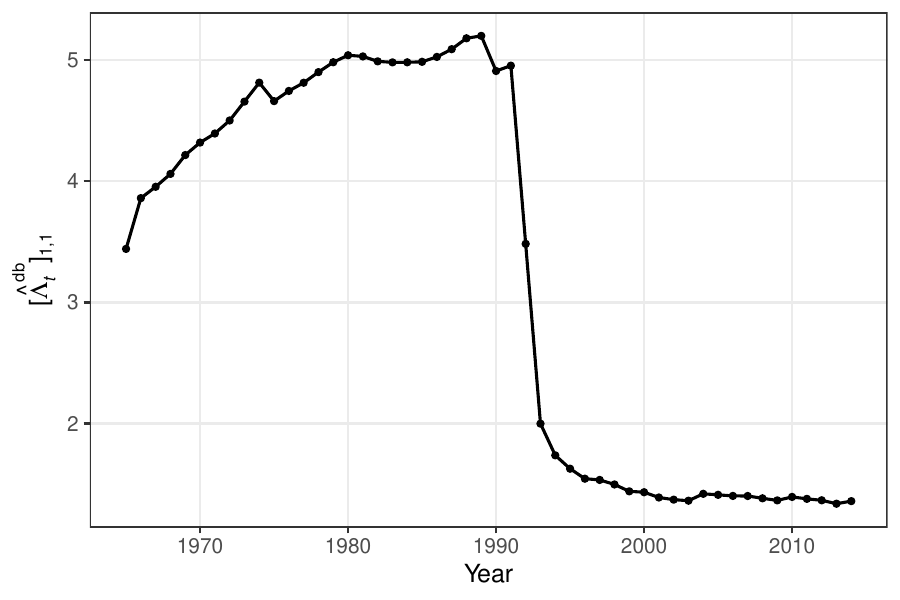}
  \caption{$\{[\widehat\LLLL_t^{\mathrm{db}}]_{1,1}\}_{t=1}^T$}
  \label{fig:core11-left}
\end{subfigure}\hfill
\begin{subfigure}[t]{0.48\textwidth}
  \centering
  \includegraphics[width=\linewidth, height=4cm]{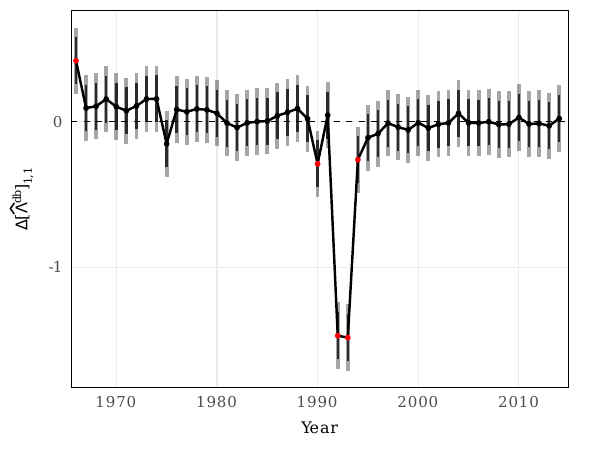}
  \caption{$\{[\widehat\LLLL_{t}^{\mathrm{db}}]_{1,1}
  -[\widehat\LLLL_{t-1}^{\mathrm{db}}]_{1,1}\}_{t=2}^{T}$}
  \label{fig:core11-right}
\end{subfigure}

\caption{
The left panel shows the estimated sequence of
$[\widehat\LLLL_t^{\mathrm{db}}]_{1,1}$ over time.
The right panel displays the corresponding first-order differences with
two $95\%$ confidence intervals at each time point: the black intervals are
the original confidence intervals, while the gray intervals are
Bonferroni-corrected intervals. The red dots indicate statistically
significant change points identified using the Bonferroni-corrected
confidence intervals.
}
\label{fig:core11}
\end{figure}

The sequence $\{\LLLL_t\}_{t=1}^T$ captures the evolution of the connection pattern of the multilayer network. Specifically, $[\LLLL_t]_{i,j}$ is the coefficient for the interaction effect between the $i$-th column of $\TTT$ and the $j$-th column of $\PH$.
Figure~\ref{fig:core11-left} shows $\{[\widehat\LLLL_t^{\mathrm{db}}]_{1,1}\}_{t=1}^T$, and it exhibits a marked structural break in the early 1990s, coinciding with the breakup of the Soviet Union. The emergence of new countries and the dissolution of others introduced substantial structural changes to the global trade network. We further test whether the structural changes are significant by the multiple testing procedure described in Section~\ref{subsection:estimation_asymp_dist_core} with $t' = t - 1$ and $\alpha = 0.05$.
To this end, we estimate the noise variance $\sigma_0^2$ by the sample variance of the entries in the tensor $[\YYY;\JJ_n,\JJ_n,\II_T] - [\widehat\SSSS;\widehat\TTT,\widehat\PH,\II_T]$. Theoretical details are provided in the supplement. 
The detected change points are $\{1966, 1971,1975, 1990,1992,1993,1994,2001\}$. 
These change points align with major historical shifts in the global economy. The cluster of years in the early 1990s (1990-1994) directly corresponds to the collapse of the Soviet Union and the economic reorientation of Eastern Europe. The change point around 2001 coincides with China's entry into the WTO, which dramatically reshaped global supply chains. Earlier shifts, like 1971 and 1975, match the end of the post-war monetary system and the oil price shocks, respectively. The year 1966 marks an early transition towards deeper regional integration and trade liberalization. 
Figure~\ref{fig:core11-right} further shows the first-order difference sequences $\{[\widehat\LLLL_{t}^{\mathrm{db}}]_{1,1} -[\widehat\LLLL_{t-1}^{\mathrm{db}}]_{1,1}\}_{t=2}^{T}$,  as well as the original and Bonferroni-corrected 95\% confidence intervals at each time point, confirming such structural changes in the early 1990s and around 1966. Figures for the other $\{[\widehat\LLLL_t^{\mathrm{db}}]_{i,j}\}_{t=1}^T$ are deferred to the Supplementary Material.

\section{Conclusions}\label{section:conclusion}


This paper develops a generalized multilayer latent space model and an
{\it unfolding and fusion} procedure that avoids computationally intensive
nonlinear tensor optimization. A central inferential finding is that
the fused connection-matrix estimator contains a non-negligible
second-order bias, even though each constituent factor estimator admits
a first-order asymptotic expansion. We derive an explicit representation
and a feasible estimator of this bias. The resulting debiased estimator
is asymptotically normal and enables valid inference on layer-specific
network structures. More broadly, our results illustrate that
computationally convenient fusion of high-dimensional factor estimators
may require second-order correction for valid inference on their
low-dimensional functionals.

One possible future direction is to explore layer sparsity and node degree heterogeneity in the proposed generalized multilayer latent space model, such as the scenario considered in \cite{ke2025optimal}.
Another direction is to investigate the dynamics on the connection matrices for dynamic networks to facilitate network prediction. For example, the matrix autoregressive model \citep{chen2021autoregressive} could be incorporated into our model setting. 
In addition, incorporating node-wise or edge-wise covariates into the model framework is also feasible, for example, as in \cite{xu2023covariate} and \cite{zhang2022joint}.
It would be interesting to test the effect of such covariates on the multilayer network structure. We leave these topics for future investigation.

\bibliographystyle{apalike}
\bibliography{ref}

\end{document}